\documentclass[twocolumn,aps,prd,longbibliography]{revtex4-1}
\RequirePackage{lineno}
\usepackage{amsmath}
\usepackage{amssymb}
\usepackage{amstext}
\usepackage{amsopn}
\usepackage{amsfonts}
\usepackage{amsxtra}
\usepackage[english]{babel}
\usepackage{amsmath}
\usepackage{graphicx}
\usepackage{float}
\usepackage{bm}
\usepackage{multirow}
\usepackage{dcolumn}

\usepackage{hyperref}
\usepackage{CJK}
\usepackage{color}
\usepackage{lineno}

\begin{document}

\title{Impact of disorder in the charge-density-wave state of Pd-intercalated ErTe$_3$ revealed by the electrodynamic response}

\author{M. Corasaniti$^{\dag}$}
\author{R. Yang$^{\dag}$$^{\ddag}$}
\affiliation{Laboratorium f\"ur Festk\"orperphysik, ETH - Z\"urich, 8093 Z\"urich, Switzerland}

\author{J.A.W. Straquadine}
\author{A. Kapitulnik}
\author{I.R. Fisher}
\affiliation{Geballe Laboratory for Advanced Materials and Department of Applied Physics, Stanford University, Stanford CA 94305, USA}
\affiliation{Stanford Institute for Materials and Energy Sciences, SLAC National Accelerator Laboratory, 2575 Sand Hill Road, Menlo Park, California 94025, USA}

\author{L. Degiorgi$^\ast$}
\affiliation{Laboratorium f\"ur Festk\"orperphysik, ETH - Z\"urich, 8093 Z\"urich, Switzerland}

\date{\today}

\begin{abstract}
It is a general notion that disorder, introduced by either chemical substitution or intercalation as well as by electron-irradiation, is detrimental to the realisation of long-range charge-density-wave (CDW) order. We study the disorder-induced suppression of the in-plane CDW orders in the two-dimensional Pd-intercalated ErTe$_3$ compositions, by exploring the real part of the optical conductivity with light polarised along the in-plane $a$ and $c$ axes. Our findings reveal an anisotropic charge dynamics with respect to both incommensurate unidirectional CDW phases of ErTe$_3$, occurring within the $ac$-plane. The anisotropic optical response gets substantially washed out with Pd-intercalation, hand-in-hand with the suppression of both CDW orders. The spectral weight analysis though advances the scenario, for which the CDW phases evolve from a (partially) depleted Fermi surface already above their critical onset temperatures. We therefore argue that the long-range CDW orders of ErTe$_3$ tend to be progressively dwarfed by Pd-intercalation, which favours the presence of short-range CDW segments for both crystallographic directions persisting in a broad temperature ($T$) interval up to the normal state, and being suggestive of precursor effects of the CDW orders as well as possibly coexisting with superconductivity at low $T$.
\end{abstract}
\maketitle

\section{Introduction}
The interplay between broken-symmetry ground states, either via competition or coexistence, belongs to the mostly explored topics of modern solid-state physics and seems to be an essential ingredient in the arena of strongly correlated materials. It goes without saying that the discovery of a charge-density-wave (CDW) phase in the middle of the pseudogap region of the high-temperature superconducting cuprates (HTC) brought into focus the intrinsic nature of those very different cooperative electronic phenomena, suggesting novel concepts such as intertwined order \cite{Fradkin2015}. However, in HTC and other materials, like the iron-based superconductors, any kind of broken-symmetry ground state also arises in the presence of significant disorder, because of the large doping required in order to destroy the phase of the parent compound (i.e., Mott insulator or magnetic). In the absence of electron-electron interactions, a metal can turn into an Anderson insulator \cite{Anderson1958} with disorder, but it can remain metallic if interactions are relevant. Therefore, disorder, intended from a broad perspective and induced either by doping or by electron-irradiation, cannot be neglected, when addressing the emergence of superconductivity from or within a CDW environment. 

It is a widespread wisdom of competing orders that the disorder-induced suppression of the CDW state may naturally lead to the increase in the superconducting critical temperature $T_c$. In a weak coupling picture, CDW suppression increases the number of carriers available for superconductivity pairing at the Fermi surface (FS), thus enhancing $T_c$. The physical scenario that in systems, where CDW competes with superconductivity, disorder promotes the latter is very significant and even extends to a strongly coupled situation as long as disorder remains weak. Though, there is a need to better scrutinise such a proposition, particularly with model systems capturing the impact of disorder but allowing to circumvent the complications of strong magnetic correlations or of a Mott insulating phase. In this respect, several old materials were revisited and several new ones were chased. For instance, members of the transition metal dichalcogenides family \cite{Li2017,Cho2018} as well as alternative materials like Cu$_x$TiSe$_2$ \cite{Morosan2006,Qian2007,Kogar2017} were intensively investigated in order to elucidate how the CDW order gets affected and conversely how superconductivity may be favoured by the presence of disorder. 

Here, we address the Pd-intercalated ErTe$_3$ (from now on also noted as Pd$_x$ErTe$_3$), which recently arose to a model family of materials serving as suitable playground for a controlled tuning of disorder \cite{Straquadine2019,Fang2019}, with the intent to trace its implications from the perspective of the charge dynamics. Our focus is specifically restricted to the optical signatures of the CDW ground state, being the Pd content $x \le$ 1.2 \% in our samples (while the onset of superconductivity clearly occurs for $x >$ 2\% \cite{Straquadine2019}). A peculiar asset of the chosen compounds resides in the fact that, unlike the parent ditelluride material where vacancies were found, the (undoped) tritelluride one is free from defects in the Te layers and the accurately tunable Pd-intercalation in ErTe$_3$ solely allows a more systematic optical investigation than previous attempts \cite{Huang2012,vacancies}. ErTe$_3$ itself originally kindled a lot of attention since it belongs to the wider series of $R$Te$_3$ ($R$ = Y, La-Nd, Sm, Gd-Tm) quasi two-dimensional metals exhibiting unidirectional (in-plane) incommensurate CDW states and harbours two successive CDW phase transitions, with critical temperatures $T_{CDW1} \simeq$ 260 K and $T_{CDW2} \simeq$ 160 K \cite{DiMasi1995,Ru2008a,Ru2008b,Fang2020}. The resistivity ($\rho_{dc}(T)$) measurements \cite{Walmsley2017} provide signatures for anisotropic transport properties at both CDW phase transitions (i.e., broad bumps overlapped to the otherwise metallic-like $\rho_{dc}(T)$ for each in-plane $a$ and $c$ axis, Fig. \ref{resistivity} in Appendix), which are then smeared and suppressed (i.e., lowering of both $T_{CDW1}$ and $T_{CDW2}$) by the Pd-intercalation, consistent with the dominant effect arising from disorder \cite{Straquadine2019}.

The absorption spectrum \cite{Dressel2002} generally gives access to the relevant energy scales (like the CDW gap(s)), and the optical spectral weight ($SW$) encountered in the charge dynamics as well as its relative temperature ($T$) evolution shed light on the reconstruction of the electronic band structure and on FS instabilities with respect to the CDW transitions \cite{Kohn1959,Gruner2000,Eiter2013}, here additionally singled out for both the $c$ and $a$ axes (Fig. \ref{resistivity}(a) in Appendix). We will impart that our present data grasp first of all an anisotropic optical response (missed in previous investigations \cite{Pfuner2010,Hu2011}), emerging between the two in-plane crystallographic axes but diminishing with disorder, and second emphasise the relevance of CDW precursor effects, the latter present already at $T > T_{CDW1}$ and evolving into a long-range CDW order at $T < T_{CDW1}$ and $T_{CDW2}$ in the pristine ($x$ = 0) and weakly Pd-intercalated Pd$_x$ErTe$_3$.  

\section{Samples and Experiment}
The samples for our optical investigation were grown after a Te self-flux method as described previously for $R$Te$_3$ ($R$ = rare-earth) \cite{Ru2006}, with the additional small amounts of Pd to the melt. The resulting crystals are orange in color, shiny, soft, and micaceous metals. 

The $T$ dependence of the optical reflectivity ($R(\omega)$) at near-normal incidence of light is measured on shiny surfaces, pertaining to the $ac$-plane (Fig. \ref{resistivity}(a) in Appendix), of an approximate size of 1$\times$2~mm$^2$ for all samples ~\cite{Dressel2002}. Measurements were always performed on freshly cleaved (thin) crystals for both polarisations of light along the $c$ and $a$ axes. Data from $\sim 30$ to $12\,000$~cm$^{-1}$, thus from the far- (FIR) up to the mid- (MIR) and then near-infrared (NIR) spectral ranges, are collected from $\sim 5$ to 300~K by using the Fourier transform infrared spectrometer (Bruker Vertex 80v). From NIR up to the ultra-violet (UV) range, i.e. 4\,000$\sim$48\,000~cm$^{-1}$, $R(\omega)$ is measured at 300 K with the PerkinElmer Lambda 950 spectrometer.

The $R(\omega)$ spectra over the broad FIR-UV spectral range are then an indispensable prerequisite in order to preform reliable Kramers-Kronig transformation for achieving the real part ($\sigma_1(\omega)$) of the broadband (longitudinal) optical conductivity \cite{Dressel2002}. We refer to the Appendix for further technical details about the experiment and for a comprehensive review of the original $R(\omega)$ data (Figs. \ref{reflectivity_0} - \ref{reflectivity_012}) and corresponding $\sigma_1(\omega)$ (Figs. \ref{opt_cond_0} - \ref{opt_cond_012}). 

\section{Results and Discussion}

\begin{figure}
\center
\includegraphics[width=8.5cm]{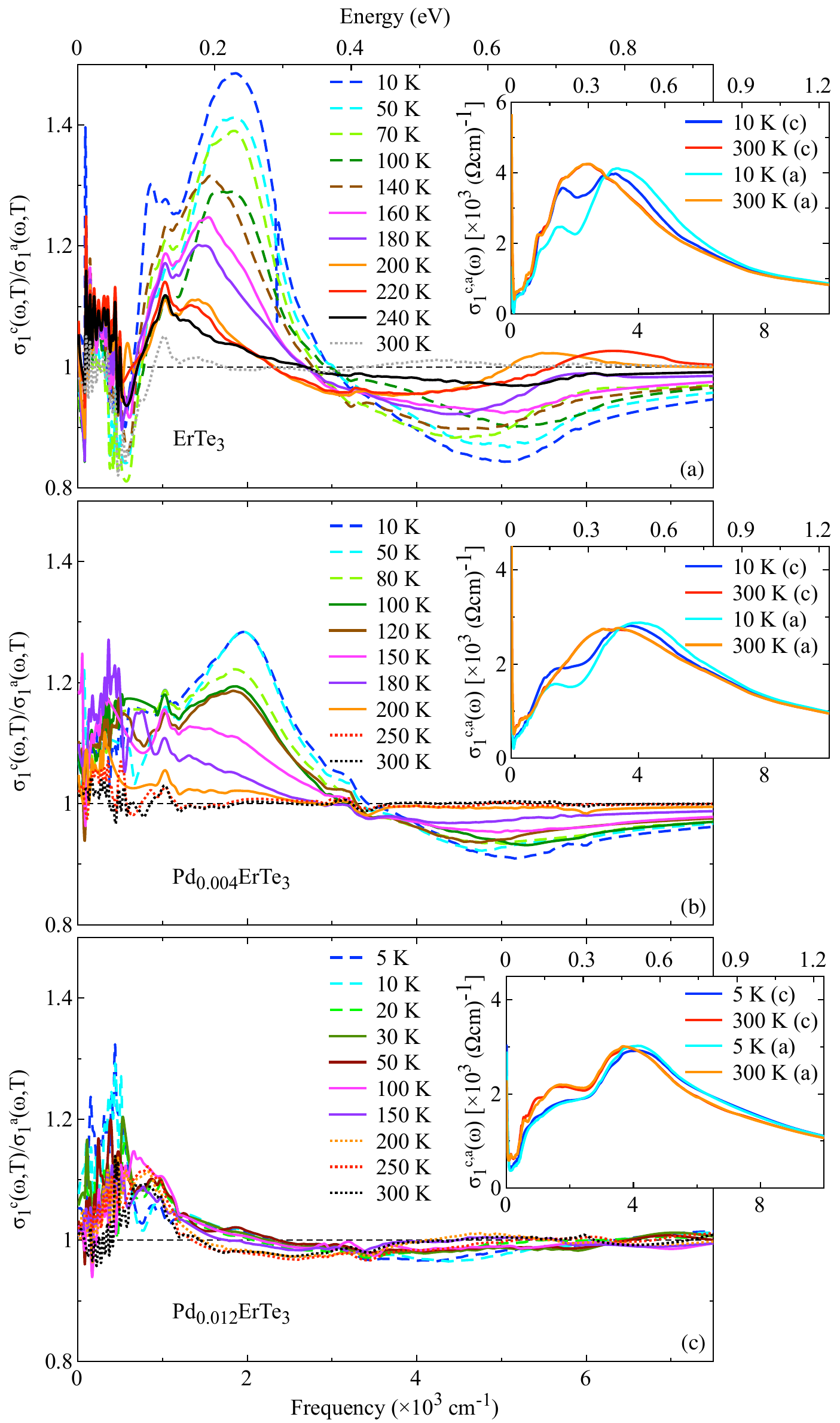}
\caption{$T$ dependence of the optical anisotropy, defined by the ratio  $\sigma_1^{c}(\omega, T)/\sigma_1^{a}(\omega, T)$, for (a) $x$ = 0, (b) $x$ = 0.004 and (c) $x$ = 0.012 Pd$_x$ErTe$_3$ from the FIR up to the MIR energy scales (1 eV = 8.06548$\times$10$^3$ cm$^{-1}$). The dotted lines refer to $T > T_{CDW1}$, steady lines to $T_{CDW2} < T < T_{CDW1}$ and dashed lines to $T < T_{CDW2}$. The insets show $\sigma_1(\omega)$ for both the $c$ and $a$ axes at 300 K and 5 or 10 K, as the highest and lowest measured $T$, respectively, covering the entire spectral range up to the visible frequencies (see Appendix for further data and details).
} 
\label{anisotropy}
\end{figure}

\begin{figure*}
\center
\includegraphics[width=17cm]{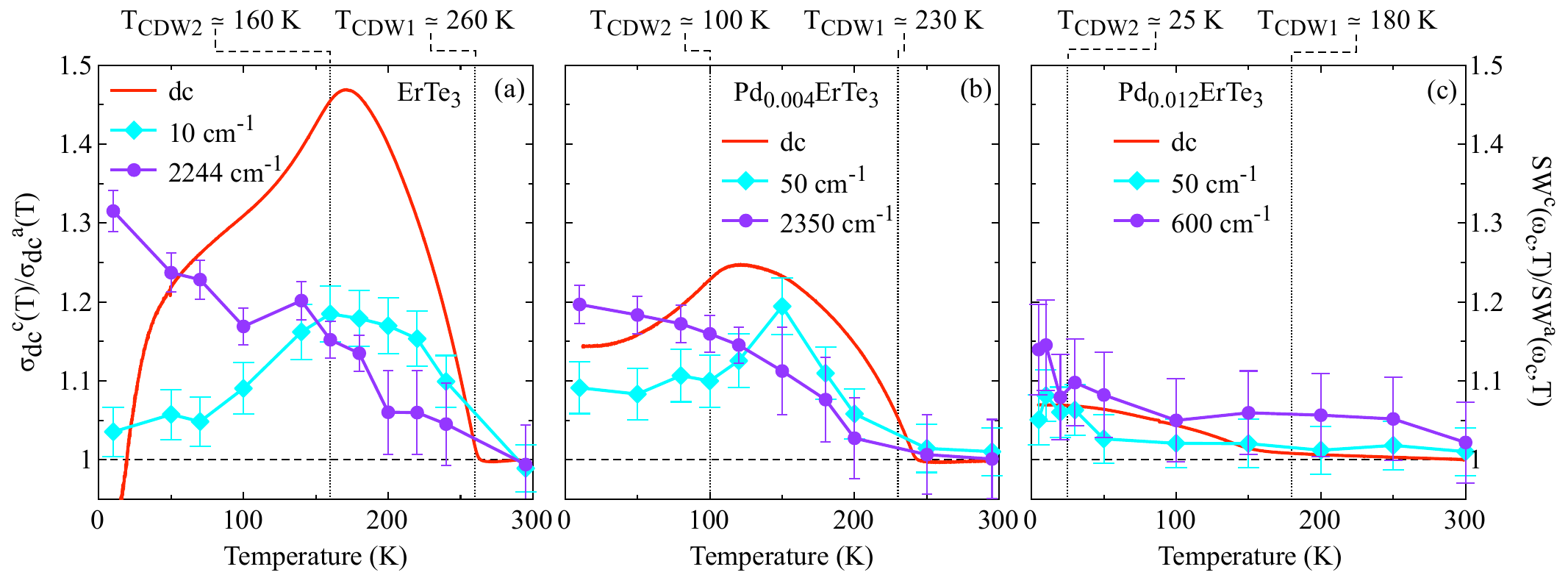}
\caption{$T$ dependence of the $dc$ transport anisotropy, defined by $\sigma_{dc}^{c}(T)/\sigma_{dc}^{a}(T)$ (with $\sigma_{dc}(T) = \frac{1}{\rho_{dc}(T)}$)  \cite{dcanisotropy}, compared to the anisotropy of the optical $SW$ given by $SW^{c}(\omega_c, T)/SW^{a}(\omega_c, T)$ (from Fig. \ref{int_SW} in Appendix) at selected cut-off frequencies $\omega_c$ \cite{SW}, representing the FIR (i.e., $\omega_c \sim$ 10-50 cm$^{-1}$, therefore significant for the $dc$ limit of the optical response) and MIR (i.e., $\omega_c \sim$ 600-2400 cm$^{-1}$) spectral range: for (a) $x$ = 0, (b) $x$ = 0.004 and (c) $x$ = 0.012 Pd$_x$ErTe$_3$. The vertical dotted lines mark $T_{CDW1}$ and $T_{CDW2}$ \cite{Straquadine2019}. See text about the origin of the error bars in $SW^{c}(\omega_c, T)/SW^{a}(\omega_c, T)$.
} 
\label{anisotropy_dc}
\end{figure*}

Introducing our results, we primarily highlight the derived optical anisotropy as a function of $T$, defined by the $\sigma_1(\omega)$ ratio between the $c$ and $a$ axes (i.e., $\sigma_1^{c}(\omega, T)/\sigma_1^{a}(\omega, T)$), shown in Fig. \ref{anisotropy} for the FIR up to the MIR spectral range. As pointed out earlier for the pristine compound, this FIR-MIR energy interval mates with the spectral range tied to the characteristic energy scales of the CDW gap(s), located at about 3000 cm$^{-1}$ \cite{Pfuner2010,Hu2011,Hu2014}. Such an anisotropy in ErTe$_3$ (Fig. \ref{anisotropy}(a)) is mainly dominated by the peak around 2000 cm$^{-1}$ and the depletion around 5000 cm$^{-1}$, both coinciding with related absorption features at equivalent energy scales in $\sigma_1(\omega)$ (inset of Fig. \ref{anisotropy}(a)). The evolution in $T$ of the intensity in $\sigma_1^{c}(\omega, T)/\sigma_1^{a}(\omega, T)$ is quite monotonous upon crossing $T_{CDW1}$ and $T_{CDW2}$. Moreover, the observed optical anisotropy progressively pales upon increasing the Pd-intercalation content $x$ (Figs. \ref{anisotropy}(b,c) and their insets). Overall, we anticipate that these findings bear testimony to an important reconstruction of the electronic band structure as well as to a substantial spillover effect of the intercalation-driven disorder with respect to both CDW transitions.

\begin{figure*}
\center
\includegraphics[width=17cm]{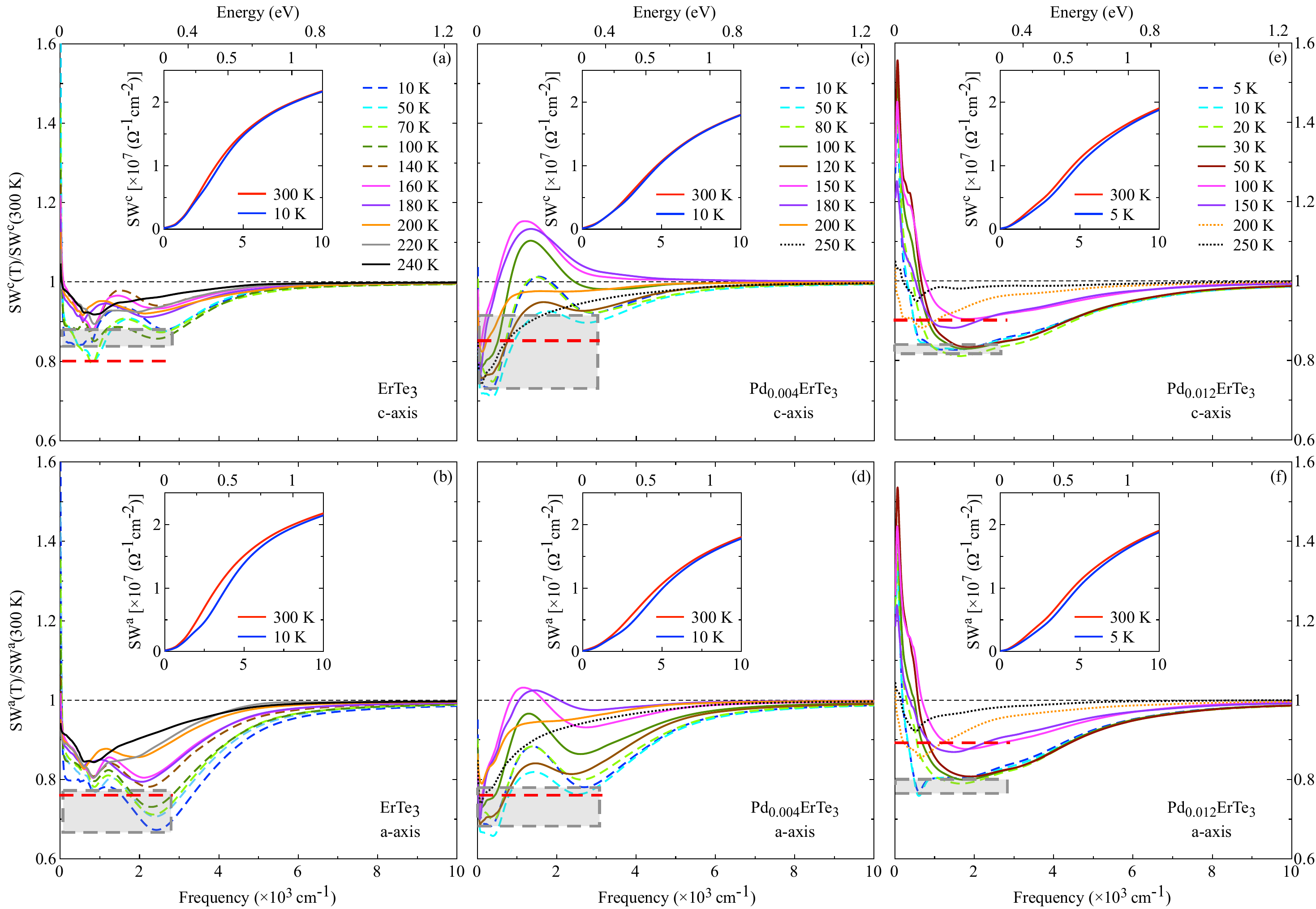}
\caption{$T$ dependence of the integrated spectral weight ratio $SW^{c,a}(T)/SW^{c,a}(300$ K) along the $a$ (lower panels) and $c$ (upper panels) axes for (a,b) $x$ = 0, (c,d) $x$ = 0.004 and (e,f) $x$ = 0.012 Pd$_x$ErTe$_3$. The dotted lines refer to $T > T_{CDW1}$, steady lines to $T_{CDW2} < T < T_{CDW1}$ and dashed lines to $T < T_{CDW2}$. The range, within which the depletion of the $SW$ ratio occurs at the lowest $T$ across the FIR-MIR energy interval, is noted by the grey areas. It roughly indicates the amount of the FS-gapping. The horizontal red-dashed lines represent an estimation \cite{Ong1977,rho} of the amount of the FS-gapping from $\rho_{dc}(T)$ \cite{Walmsley2017,Straquadine2019,dcanisotropy}. The insets show the integrated $SW^{c,a}(T)$ at 5 or 10 and 300 K (in units of $\Omega^{-1}$cm$^{-2}$ \cite{SW}) for each crystallographic axis. Main panels and insets cover the energy spectral range from zero up to $\omega_c \sim$ 10$^4$ cm$^{-1}$ (1 eV = 8.06548$\times$10$^3$ cm$^{-1}$), since above this energy scale $SW^{c,a}(T)$ is fully recovered and conserved at any $T$ \cite{Dressel2002}.
} 
\label{SW_ratio}
\end{figure*}

Figure \ref{anisotropy_dc} anew underscores the anisotropy of the optical response, depicted here by the ratio of the integrated spectral weight $SW^{c}(\omega_c, T)/SW^{a}(\omega_c, T)$ (Fig. \ref{int_SW} in Appendix) at selected cut-off frequencies $\omega_c$ \cite{SW}. We choose two $\omega_c$; at FIR energy scales (i.e., 10-50 cm$^{-1}$), pertaining to the metallic Drude component of $\sigma_1(\omega)$ (see Figs. \ref{SW_fit0}(b,d,f) and \ref{SW_fit90}(b,d,f) in Appendix), and at MIR ones (i.e., 600-2400 cm$^{-1}$), denoting the energy interval with the occurrence of the optical anisotropy affiliated to the interband transitions (Fig. \ref{anisotropy}). We exert the criterion to set $\omega_c$ at MIR frequencies in coincidence with the peak of $SW^{c}(\omega_c, T)/SW^{a}(\omega_c, T)$ (Fig. \ref{int_SW} in Appendix). We remark that $\omega_c$ is uniquely identified and it is virtually constant at $T \le$ 150 - 200 K, while $\omega_c$ turns out to be less precisely defined at higher $T$. This propagates to the uncertainties in the estimation of the anisotropic $SW^{c}(\omega_c, T)/SW^{a}(\omega_c, T)$ and it is accounted for by the enhanced error bars upon increasing $T$ in Fig. \ref{anisotropy_dc}.

$SW^{c}(\omega_c, T)/SW^{a}(\omega_c, T)$ is contrasted with the $dc$ transport anisotropy, represented by the ratio of the $dc$ conductivity along the $c$ and $a$ axes (i.e., $\sigma_{dc}^{c}(T)/\sigma_{dc}^{a}(T)$) \cite{dcanisotropy}. For $\omega_c$ at FIR energy scales $SW^{c}(\omega_c, T)/SW^{a}(\omega_c, T)$ mimics the anisotropy of the $dc$ conductivity. On the contrary, for $\omega_c$ at MIR energy scales the optical anisotropy increases gradually, with a change of slope between $T_{CDW1}$ and $T_{CDW2}$ being barely perceptible upon increasing the Pd-intercalation. This is already a signature that the conduction bands crossing the Fermi energy ($E_F$) directly reflect the $dc$ anisotropy as a function of $T$, while the implications of the CDW states on the electronic band structure at energy scales far away from $E_F$ are somehow more distinct and the resulting (optical) anisotropy is not necessarily copying with the $dc$ transport-like behaviour across $T_{CDW1}$ and $T_{CDW2}$. It is also worth noting that the anisotropy of all quantities gets considerably reduced upon increasing the Pd-intercalation concentration $x$.

We stick to the model independent analysis based on the integrated $SW$ \cite{SW} and consider, complementary to Fig. \ref{anisotropy_dc}, the $SW$ ratio with respect to 300 K, given by $SW^{c,a}(T)/SW^{c,a}(300$ K) and shown as a function of $T$ in Fig. \ref{SW_ratio} for the $a$ (lower panels) and $c$ (upper panels) axis. The $SW$ ratio at $T < $ 300 K clearly emphasises a two-fold $SW$ reallocation to low (with $SW$ ratio above 1) and above all to high (with $SW$ ratio below 1) energies (see also Appendix for more insights), which gets more pronounced upon decreasing $T$. This seems to be true for all Pd$_x$ErTe$_3$. Such a behaviour is anisotropic as well, since the depletion in $SW^{c,a}(T)/SW^{c,a}(300$ K) below 1 is clearly stronger along the $a$ than $c$ axis (at least for $x$ = 0 and 0.004); an aspect which may be already recognised directly from the integrated $SW$ (insets of Fig. \ref{SW_ratio}), being it at 5 or 10 K in the MIR spectral range decidedly more depleted along the $a$ than $c$ axis with respect to 300 K. The accumulation of $SW$ at low energies for both polarisations occurs in a rather small FIR spectral range, which narrows as well upon decreasing $T$ (for $x$ = 0 and 0.004 Pd$_x$ErTe$_3$) or at least at low $T$ (for $x$ = 0.012 Pd$_x$ErTe$_3$). This narrowing may be ascribed to the suppression of scattering channels for the itinerant charge carriers in the CDW ground state(s) (Figs. \ref{Drude_param}(a,c,e) in Appendix). On the other hand, the width at any $T$ of that low energy $SW$ accumulation becomes visibly larger for the highest Pd-intercalation at $x$ = 0.012 (Figs. \ref{SW_ratio}(e,f)), being this a direct fingerprint of the enhanced Drude scattering rate along both polarisation directions (Fig. \ref{Drude_param}(e) in Appendix) as consequence of the Pd-intercalation-induced disorder. 

Figure \ref{SW_ratio} also tells us that a smaller fraction of FS gets gapped by the CDW transitions \cite{Kohn1959,Gruner2000} upon increasing the Pd-intercalation concentration $x$. Indeed, the depletion below 1 of the $SW$ ratio itself is a rough measure of that fraction \cite{Corasaniti2019}, which is of the order of 10 to 30 \% (graphically sketched by the grey areas in Fig. \ref{SW_ratio}). Particularly along the $a$ axis, for which the anomalies in the $dc$ resistivity are remarkably more pronounced than along the $c$ axis (Figs. \ref{resistivity}(b-d) in Appendix), the effective FS-gapping tends to shrink with increasing Pd-intercalation. We might ponder that the amount of the FS-gapping along the $c$ axis is less influenced by the Pd-intercalation than along the $a$ axis for both (primary and secondary) CDW transitions. The relationship and correspondence between the FS-erosion and the anomaly in $\rho_{dc}$ were already advanced in the past \cite{Ong1977} in connection with the CDW as well as SDW transitions in two-dimensional materials (for which notoriously the FS-nesting is not as perfect as in one dimension  \cite{Kohn1959,Gruner2000,Eiter2013}). Independent from the origin of the FS-gapping or depletion upon which we will return later, we find a reasonable agreement within the same bulk values of its estimation from the $SW$ ratio (grey areas in Fig. \ref{SW_ratio}) and the $dc$ transport data (horizontal red-dashed lines in Fig. \ref{SW_ratio} \cite{rho}) of Pd$_x$ErTe$_3$. We caution however that the presence of two CDW phase transitions at $T_{CDW1}$ and $T_{CDW2}$ and of the corresponding double anomaly in $\rho_{dc}(T)$ may configure such a comparison as disputable. Nonetheless, this appraisal looks a posteriori rather robust. The discrepancy between the two analysis at the $x$ = 0.012 Pd-intercalation (Figs. \ref{SW_ratio}(e,f)) is explained by the use of $\rho_{dc}(T)$ for the $x$ = 0.019 one (Fig. \ref{resistivity}(d) in Appendix), which consistently dissimulates a smaller FS-gapping. As a note on the side, the frequency of the deepest minimum in $SW^{c,a}(T)/SW^{c,a}(300$ K) may be used as an indicative, notwithstanding rough measure of the CDW gap(s) \cite{Pfuner2010,Hu2011,Hu2014}. We observe a broadening of the minimum in the $SW$ ratio between the $x$ = 0 and 0.012 Pd-intercalations, testifying an amplified distribution of the gap energy scales and respectively an additional manifestation of disorder besides its generated reduction of the effective amount of the FS-gapping.

We finally turn our attention to the $T$ dependence of the $SW$ ratio and its consequences towards a possible microscopic model for the interplay between disorder and CDW phase transitions. Foremost, it is worth knowing that a depletion of the $SW$ ratio occurring over an ample $T$ interval, even extending well above the nominal CDW transition temperatures, was also recently observed in ZrTe$_{3-x}$Se$_x$ \cite{Chinotti2018}. In fact, the $SW$ ratio in Fig. \ref{SW_ratio} is consistently depleted in the MIR spectral range at all $T$ with respect to 300 K along both polarisation directions. Admittedly, for ErTe$_3$ one may argue that the depletion of the $SW$ ratio just happens for $T$ already below $T_{CDW1}$, since the relatively high $T_{CDW1}$ and our $T$-grid do not allow any firm claim for the optical response of the so-called normal state. However, for the two Pd-intercalated ErTe$_3$ compositions studied here (Figs. \ref{SW_ratio}(c-f)) it is a well-established experimental fact that the $SW$ removal at MIR frequencies does already happen in the normal state, so at $T$ definitely above the highest CDW transition at $T_{CDW1}$. Interestingly enough, at $T_{CDW2} < T < T_{CDW1}$ an intriguing local recovering of $SW$ is prominently observed in a narrow energy interval around 1000 cm$^{-1}$ for the $x$ = 0.004 Pd-intercalation (Figs. \ref{SW_ratio}(c,d)), affecting the low-frequency shoulder of the MIR peak prior a renewed $SW$ depletion and full recovery at higher energy scales. This latter facet is just adumbrated in the pristine compound ($x$ = 0, Figs. \ref{SW_ratio}(a,b)) and is not at all detected in the spectra of the $x$ = 0.012 Pd-intercalation (Figs. \ref{SW_ratio}(e,f)). While beyond the immediate scope of this work, we venture a guess that this not-straightforward $SW$ reshuffling in selected spectral ranges underlines the complex evolution of the electronic band structure and its interplay with impurity bands in the intermediate $T$ regime between the two CDW transitions. We refer to the discussion around Figs. \ref{SW_fit0} and \ref{SW_fit90} in Appendix for an alternative analysis of the $SW$ redistribution, pinned down to specific energy intervals.

\section{Conclusions}

The globally emerging picture is tightly bound to the presence of a pseudogap in the electronic excitation spectrum of Pd$_x$ErTe$_3$ in their normal-state. Such pseudogaps reveal the tendency towards the formation of short-range order, uncorrelated CDW segments forming already at high $T$ (i.e., in the normal state) and crossing over to a coherent CDW condensate at low $T$. We conjecture that the Pd-intercalation may then perturb the long-range phase coherence of the pristine material by inducing patches of CDW condensates without specific and/or well defined orientations with respect to the crystallographic structure, as similarly proposed elsewhere \cite{Zhu2016}. This destroys the bidirectional nature of the CDW transitions, waving the characteristic anisotropy in several physical properties between the crystallographic axes but leading yet to the partial depletion of FS for any Pd-intercalation (Fig. \ref{SW_ratio}).

Our optical results therefore convey the relevance of precursors in the formation of the CDW condensate, which certainly are not surprising for genuine CDW materials \cite{Schwartz1995,Perucchi2004,Gruner2000}. In the latter materials, precursor effects predominantly originate from fluctuations of the order parameter, as originally proposed in the seminal work of Ref. \onlinecite{Lee1973}. Additionally, our findings directly map the impact of the intercalation-induced disorder, which concerns both the $dc$ transport data as well as the optical outcome at any FIR-MIR energy scales (Fig. \ref{anisotropy_dc}). In this context, alternative avenues may be considered for the interplay between CDW transition and disorder itself. It is worth recalling the setting based on the excitonic insulator dynamics \cite{Kohn1967,Kidd2002}, which was advanced as major ingredient for the (Overhauser-type) CDW state in TiSe$_2$ \cite{Li2007}. In fact, Cu intercalation in TiSe$_2$ suppresses such excitonic correlations leaving nonetheless the electron-phonon interaction less affected and thus allowing the presence of CDW incommensurations \cite{Qian2007,Kogar2017}. Whether this is likely pertinent for our materials calls for further investigations. As a matter of fact, contrary to the semimetallic nature of the CDW state in 1$T$-TiSe$_2$ the shape of FS in ErTe$_3$, evinced from the angle-resolved-photoemission-spectroscopy \cite{Moore2010}, principally tends to favour the creation of CDW gaps by perpendicular FS nesting vectors. Raman data question though the nesting as unique driving mechanism for the CDW transition; instead of a purely electronic instability the CDW ordering vector could be determined by a lattice distortion driven by some other mechanism exploiting the role of the electron-phonon coupling in the spirit of a so-called focusing effect \cite{Eiter2013}. Generally, solving the dichotomy about the repercussion of disorder on the CDW collective state would be also instrumental towards superconductivity, as we now cursorily flash.

\begin{figure*}
\center
\includegraphics[width=17cm]{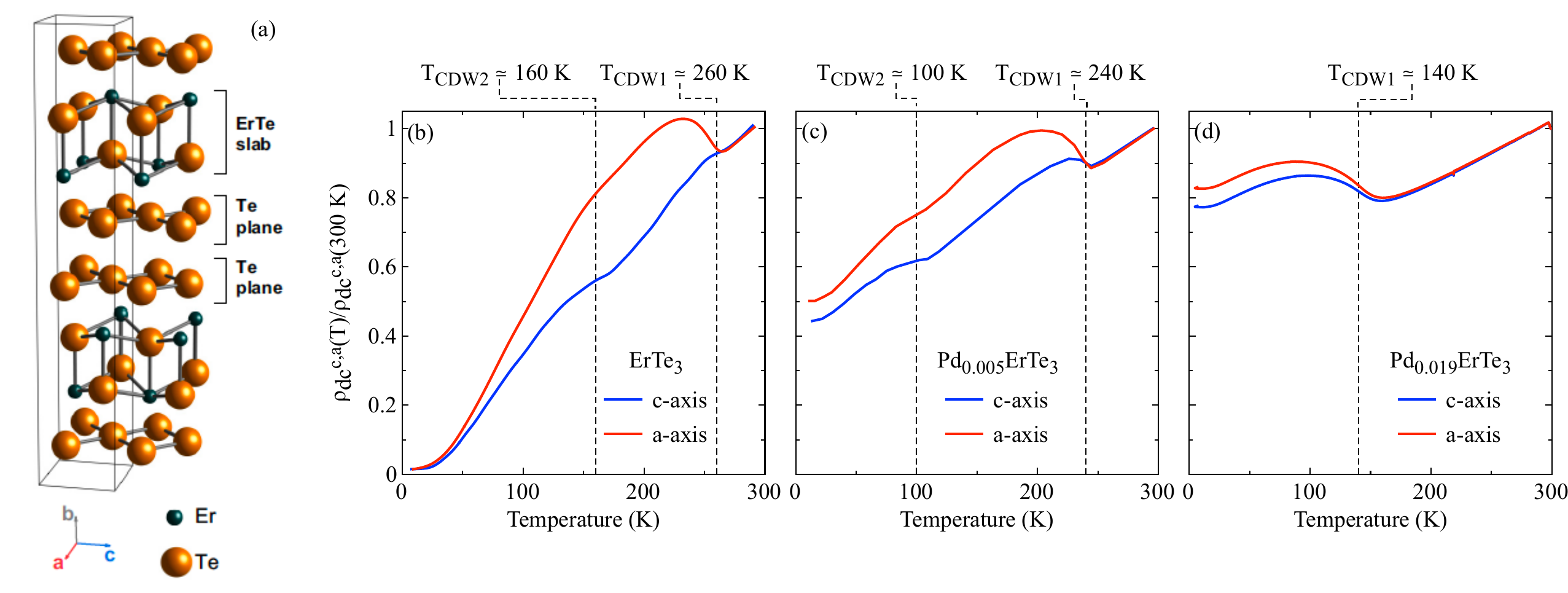}
\caption{(a) The original crystal structure of ErTe$_3$ with the relevant axes is reproduced from Ref. \onlinecite{Straquadine2019}. (b-d) $T$ dependence of the $dc$ resistivity ($\rho_{dc}(T)$) normalised at 300 K along both the $a$ and $c$ axes for (b) $x$ = 0, (c) $x$ = 0.005 and (d) $x$ = 0.019 Pd$_x$ErTe$_3$, reproduced from Refs. \onlinecite{Straquadine2019} and \onlinecite{Walmsley2017}. The vertical dashed lines mark $T_{CDW1}$ and $T_{CDW2}$. 
}
\label{resistivity}
\end{figure*}

As future outlook, it remains indeed to be seen how the gradual suppression of the CDW state and of its anisotropic character with disorder, as disclosed by our optical data, paves the way to the onset of superconductivity upon Pd-intercalation. Since $T_c$ in Pd$_x$ErTe$_3$ typically does not exceed 3 K, our optical tool must be complemented with resonant-cavity investigations in order to achieve the expected, relevant energy scales of the superconducting state. Nonetheless and even though superconductivity appears at Pd-intercalations $x >$ 2\% \cite{Straquadine2019}, upon extrapolating our results for $x \le$ 1.2\% we might be tempted to suggest that electrons involved in the superconducting pairing instead of competing for the same portions of FS do actually coexist with those of the CDW segments \cite{Chinotti2018,Fang2020} and that superconductivity generically bunches out from a pseudogap-like phase, as for HTC \cite{Fradkin2015}. The suppression of the electronic anisotropy upon Pd-intercalation in ErTe$_3$ concomitant with the onset of superconductivity may equally resemble the situation at the intertwined nematic, CDW and superconducting order, lately considered to be an ubiquitous feature in unconventional superconductors (like in iron-based and Kagome materials) \cite{Kasahara2012,Nie2022,Zheng2022}. At last, enlightening how our results could be compatible with the suggestive Bragg glass phase at weak disorder (i.e., the regime addressed here), as alluded by the scanning tunneling microscopy data \cite{Fang2019}, still needs to be addressed and would furthermore profit from the support of dedicated theoretical work.\\

\section*{Acknowledgements}
Work at Stanford was supported by the Department of Energy, Office of Basic Energy Sciences, under contract DE-AC02-76SF00515. \\

$^{\dag}$ Authors M.C. and R.Y. contributed equally to the work.\\

$^{\ddag}$ Present affiliation: Key Laboratory of Quantum Materials and Devices of Ministry of Education, School of Physics, Southeast University, Nanjing 211189, China\\

$^\ast$ Correspondence and requests for materials should be addressed to: 
L. Degiorgi, Laboratorium f\"ur Festk\"orperphysik, ETH - Z\"urich, 8093 Z\"urich, Switzerland; 
email: degiorgi@solid.phys.ethz.ch.

\section*{Appendix}

\subsection{Samples, experimental techniques and data}

\subsubsection{Samples characterisation and crystal structure}
The chemical analysis of our samples was performed in a JEOL JXA-8230 SuperProbe electron microprobe system, calibrated to ErTe$_3$ and PdTe$_2$ secondary standards \cite{Straquadine2019}. The microprobe analysis showed that $\sim$ 12\% of the Pd present in the melt incorporated into the crystals during growth and that the crystal composition is uniform to within experimental error at different spots on a crystal surface and between crystals grown in the same batch. Plate-shaped crystals ($b$ axis normal to the $ac$-plane, Fig. \ref{resistivity}(a)) 1-3 mm across were routinely produced. The crystal plate area remained fairly constant, but the resulting thicknesses decreased as Pd concentration ($x$) increased, from several hundred microns for $x$ = 0 to approximately 50 microns for $x$ = 0.05 Pd-intercalated ErTe$_3$. This offers indirect evidence that Pd atoms act as intercalants between the Te planes, in that their presence tends to disrupt and slow the rate of growth in this direction \cite{Straquadine2019}.  

The crystal structure (Fig. \ref{resistivity}(a)) is formed of alternating puckered $R$Te slabs with bilayers of approximately square nets of Te atoms. The presence of a glide plane in the stacking of these layers creates a 0.05\% \cite{Ru2008b} difference between the in-plane $a$ and $c$ axes lattice parameters at 300 K and biases the primary CDW transition to order along the $c$ axis. The space group of $R$Te$_3$ is nominally orthorhombic ($Cmcm$), even though evidences for effectively fourfold symmetric electronic properties were pointed out in the literature \cite{Straquadine2019}.

\subsubsection{$dc$ transport characterisation}
Measurements of the $T$ dependence of the resistance were performed in a Janis Supertran-VP continuous flow cryostat. The resistivity in the $ac$-plane (Fig. \ref{resistivity}(a)) was measured on thin rectangular crystals which had been cut with a scalpel and cleaved to expose a clean surface immediately before contacting. Crystals were cut such that current flows along the (101) axis, and contacts were attached to the surface in the transverse geometry \cite{Walmsley2017}. In this geometry, the sum of the resistivity components along the crystal axes $\rho_{dc}^a + \rho_{dc}^c$ and the in-plane resistivity anisotropy $\rho_{dc}^a - \rho_{dc}^c$ are measured simultaneously within the same crystal \cite{Walmsley2017,Straquadine2019}. The extracted $\rho_{dc}^a$ and $\rho_{dc}^c$ are displayed in Figs. \ref{resistivity}(b-d) for compositions coinciding or very close to the Pd-intercalations investigated in our work. One can easily recognise the two consecutive CDW transitions (at least at low Pd-intercalations) with bump-like features and/or changes of slope at $T_{CDW1}$ and $T_{CDW2}$ (dashed lines in Figs. \ref{resistivity}(b-d)), which lower with increasing Pd-intercalation. The $dc$ anisotropy, shown in Fig. 2, derives from these data in Figs. \ref{resistivity}(b-d), thus comparing our samples with $x$ = 0.004 and 0.012 Pd-intercalation to the $x$ = 0.005 and 0.019 compositions, respectively.

It has been broadly established that $T_{CDW1}$ marks the onset of the primary CDW order, while $T_{CDW2}$ of the second, orthogonal CDW component. Thus, despite the nearly tetragonal symmetry of the crystal, the phase at $T_{CDW2} < T < T_{CDW1}$ has unidirectional CDW order, while the low-temperature phase is bidirectional, but with generally inequivalent strengths. The two CDW transitions in ErTe$_3$ are associated with nesting wave vectors $q_{CDW1} \simeq 0.7c^*$ and $q_{CDW2} \simeq 0.68a^*$ (with $a^* = 2\pi/a_0$ and $c^* = 2\pi/c_0$, $a_0$ and $c_0$ are the respective in-plane lattice constants) \cite{Fang2020}. 

\subsubsection{Measured optical reflectivity}

\begin{figure}
\centerline{\includegraphics[width=1\columnwidth]{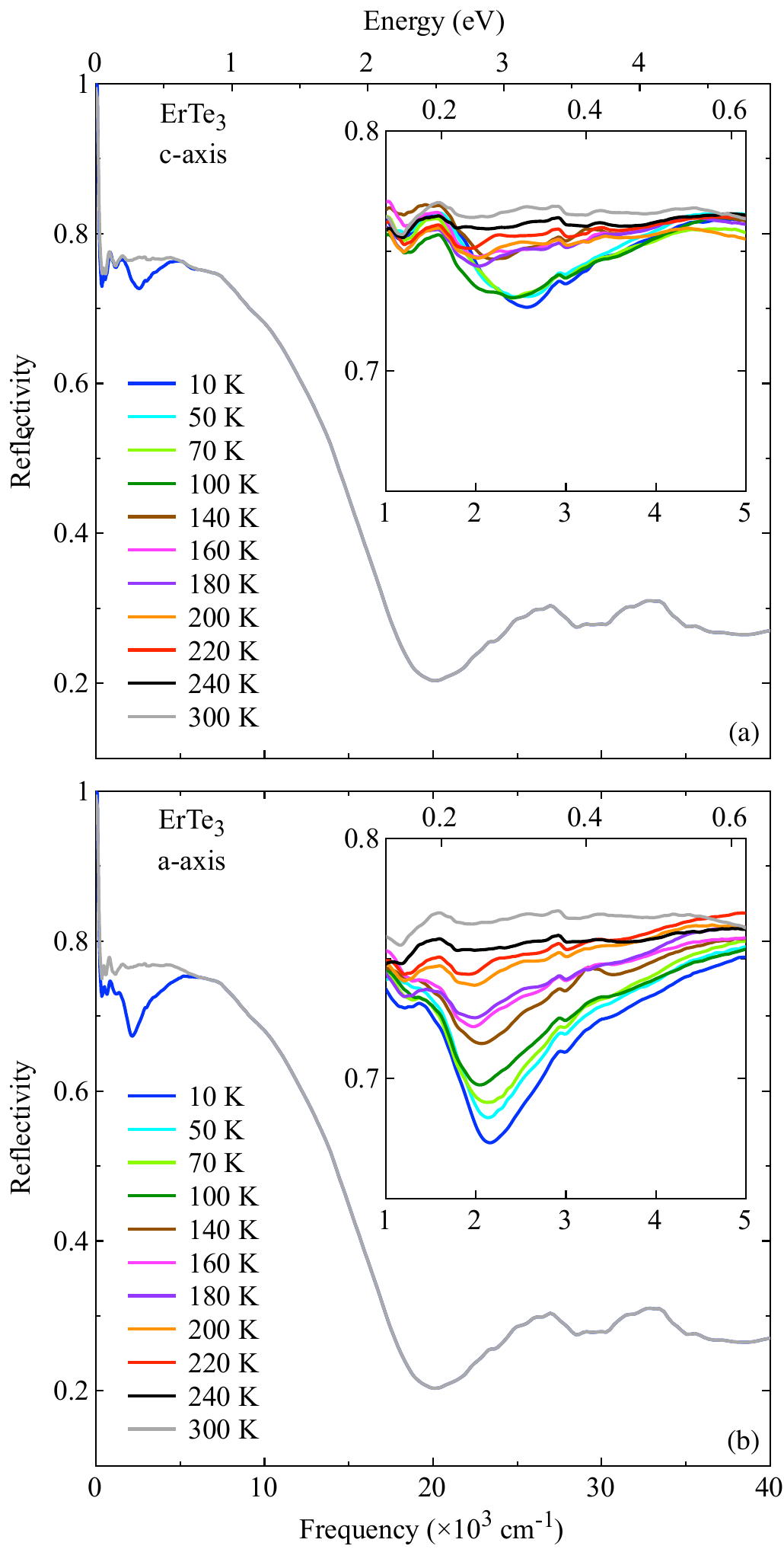}}
\caption{$R(\omega)$ along the $c$ (a) and $a$ (b) axes (Fig. \ref{resistivity}(a)) of ErTe$_3$ at 300 and 10 K (1 eV = 8.06548$\times$10$^3$ cm$^{-1}$), emphasising the plasma edge feature and the high frequency spectrum. The insets show the corresponding $T$ dependence at MIR energy scales. 
}
\label{reflectivity_0}
\end{figure}

\begin{figure}[h]
\centerline{\includegraphics[width=1\columnwidth]{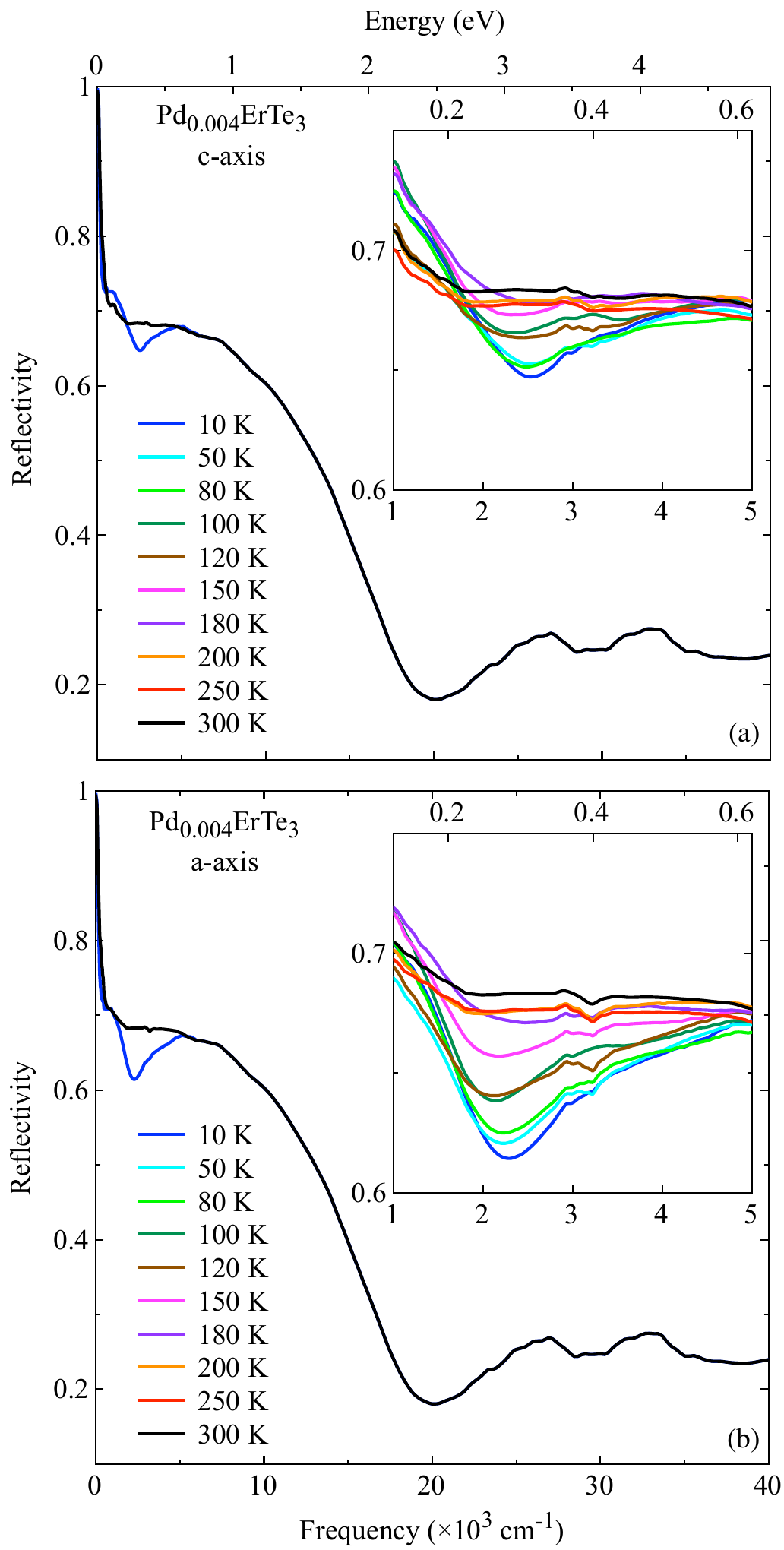}}
\caption{$R(\omega)$ along the $c$ (a) and $a$ (b) axes (Fig. \ref{resistivity}(a)) of $x$ = 0.004 Pd$_x$ErTe$_3$ at 300 and 10 K (1 eV = 8.06548$\times$10$^3$ cm$^{-1}$), emphasising the plasma edge feature and the high frequency spectrum. The insets show the corresponding $T$ dependence at MIR energy scales.
}
\label{reflectivity_004}
\end{figure}

\begin{figure}[h]
\centerline{\includegraphics[width=1\columnwidth]{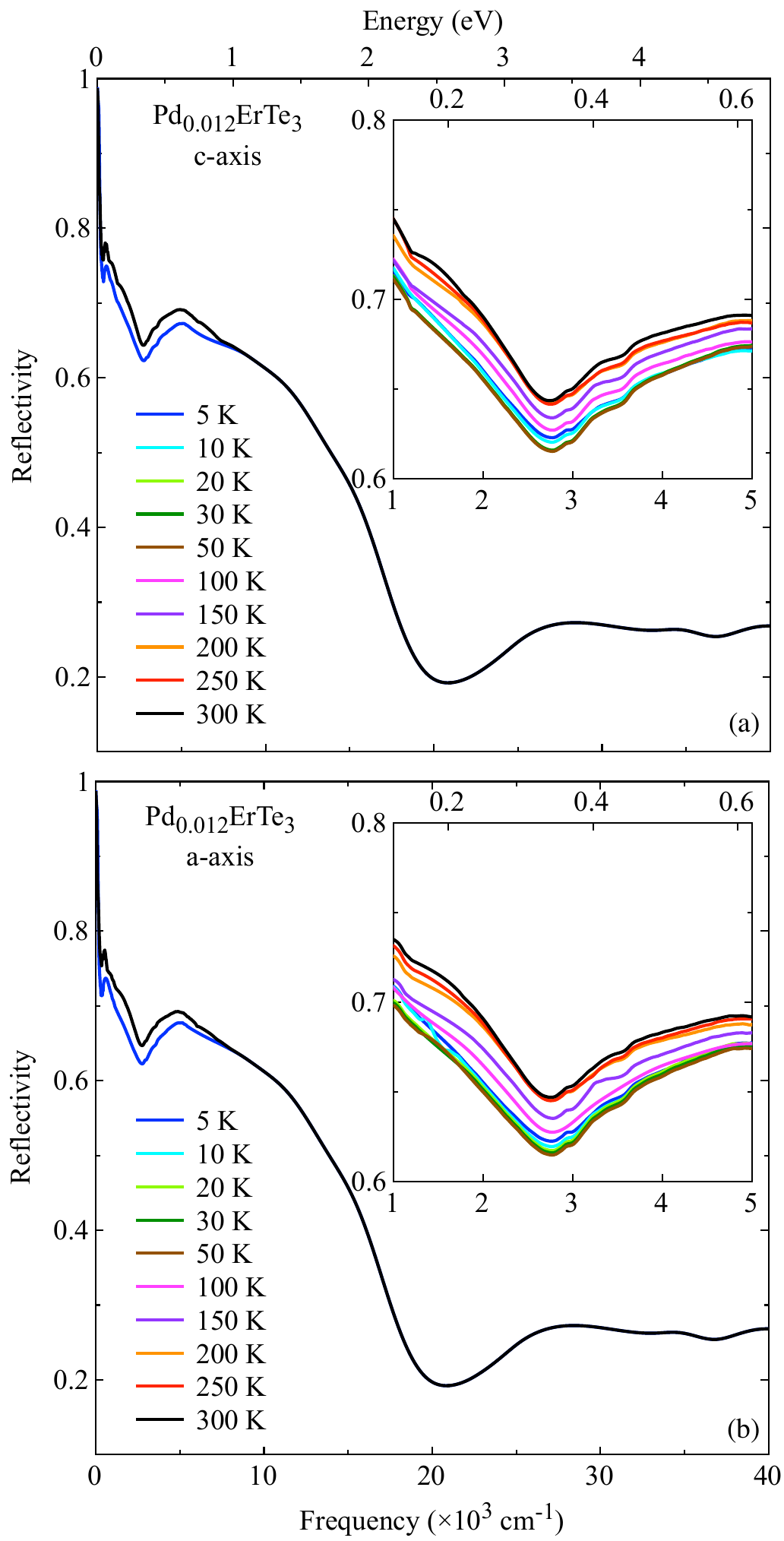}}
\caption{$R(\omega)$ along the $c$ (a) and $a$ (b) axes (Fig. \ref{resistivity}(a)) of $x$ = 0.012 Pd$_x$ErTe$_3$ at 300 and 5 K (1 eV = 8.06548$\times$10$^3$ cm$^{-1}$), emphasising the plasma edge feature and the high frequency spectrum. The insets show the corresponding $T$ dependence at MIR energy scales.
}
\label{reflectivity_012}
\end{figure}

\begin{figure*}
\center
\includegraphics[width=17cm]{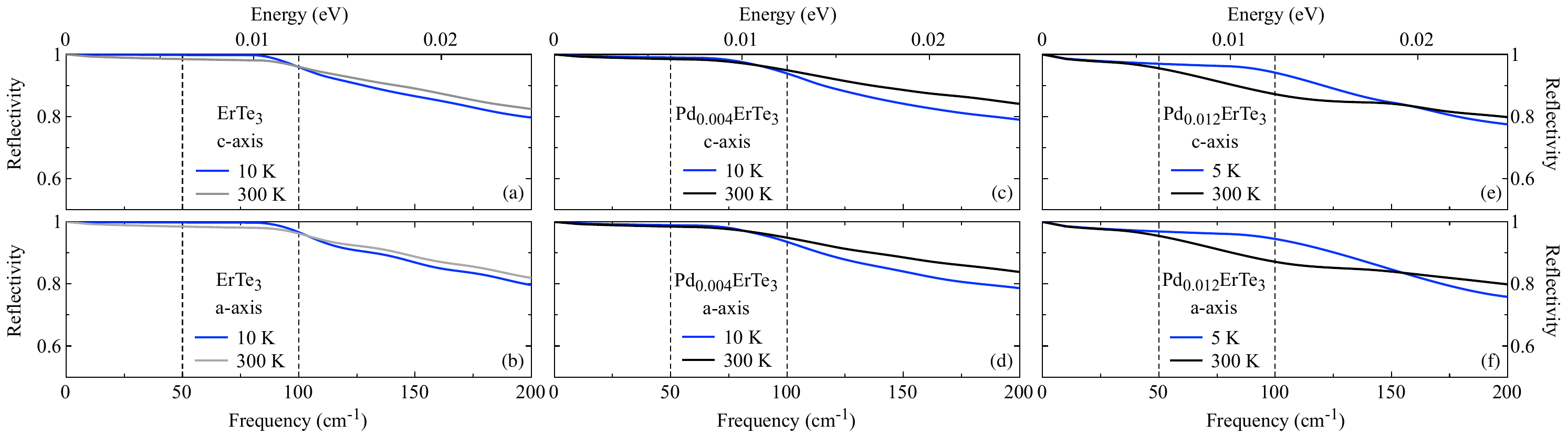}
\caption{Low frequency $R(\omega)$ at 300 and 5 or 10 K for both the $a$ and $c$ axes (Fig. \ref{resistivity}(a)) at the smooth crossover between the measured data in FIR and the HR-extrapolation: (a,b) $x$ = 0, (c,d) $x$ = 0.004 and (e,f) $x$ = 0.012 Pd$_x$ErTe$_3$. The vertical dashed lines indicate the energy interval between 50 and 100 cm$^{-1}$, where the numerical interpolation of data and HR-extrapolation takes place.
}
\label{HR}
\end{figure*}

Figures \ref{reflectivity_0}-\ref{reflectivity_012} display the measured $T$ dependence of the in-plane optical reflectivity ($R(\omega)$) for all investigated compounds along the $c$ and $a$ axes (Fig. \ref{resistivity}(a)). In all data sets, we recognise the typical metallic behaviour with a $T$-independent plasma edge (i.e., sharp increase of $R(\omega)$ upon decreasing frequency), having its onset at approximately 2$\times$10$^4$ cm$^{-1}$. While the $T$ dependence of $R(\omega)$ mainly occurs at MIR frequencies for all Pd-intercalations (insets in Figs. \ref{reflectivity_0}-\ref{reflectivity_012}), the anisotropy of the optical response appears in an obvious fashion at low $x$ but it is quite negligible for the $x$ = 0.012 Pd-intercalation. There is a fair agreement with previously collected (unpolarised) data \cite{Pfuner2010,Hu2011}, even though the overall $R(\omega)$ magnitude is here sensibly lower. This might be due to some scattering because of not perfectly flat surfaces after the fresh-cleaving, which was performed prior to each measurement. This should not impede a trusty outcome of our analysis, since we are mainly interested to quantities representing either the optical anisotropy or the relative $T$ dependence, thus based on their ratio between both polarisations or on normalised ones with respect to 300 K for each axis (Figs. 1 to 3), respectively. We also note that these data as well as our earlier unpolarised ones in ErTe$_3$ \cite{Pfuner2010} do not reveal a second deep in $R(\omega)$ around 800 cm$^{-1}$ at $T << T_{CDW2}$ either (also true for Pd-intercalated compounds), in contrast to previous claims \cite{Hu2011}. We tend to exclude major sample issues related to their growth procedure or their degradation but speculate that, if real, the resulting additional absorption feature in Ref. \onlinecite{Hu2011}, yet distinct from our low frequency shoulder at the MIR peak (see insets of Fig. 1 and below Figs. \ref{opt_cond_0}-\ref{opt_cond_012}), might be overcast by the metallic contribution in our samples.

\subsubsection{Optical conductivity and integrated spectral weight}
The real part ($\sigma_1(\omega)$) of the optical conductivity is then achieved via the Kramers-Kronig transformation of the measured $R(\omega)$.
To this goal, appropriate extrapolations of $R(\omega)$ are performed for $\omega\rightarrow 0$ and $\infty$.
Below the lowest measured frequency, the Hagen-Rubens (HR) relation $[R(\omega) = 1 - 2\sqrt{\frac{\omega}{\sigma_{dc}}}]$ for a metal is used. The values for the \emph{dc} conductivity ($\sigma_{dc}$) satisfy the relative $T$ dependence of $\rho_{dc}(T)$ (Figs. \ref{resistivity}(b-d)). The interpolation between the HR extrapolation and the measured data in FIR, performed within an energy interval between 50 and 100 cm$^{-1}$, follows a standard, common procedure and allows to smoothly connect them, as shown explicitly in Fig. \ref{HR}. Above the highest measured frequency $R(\omega)$ is first assumed to be constant up to 7~eV and above $R(\omega)$ is set to be consistent with the free-electron response ($\propto\omega^{-4}$)~\cite{Dressel2002}.

\begin{figure}
\centerline{\includegraphics[width=1\columnwidth]{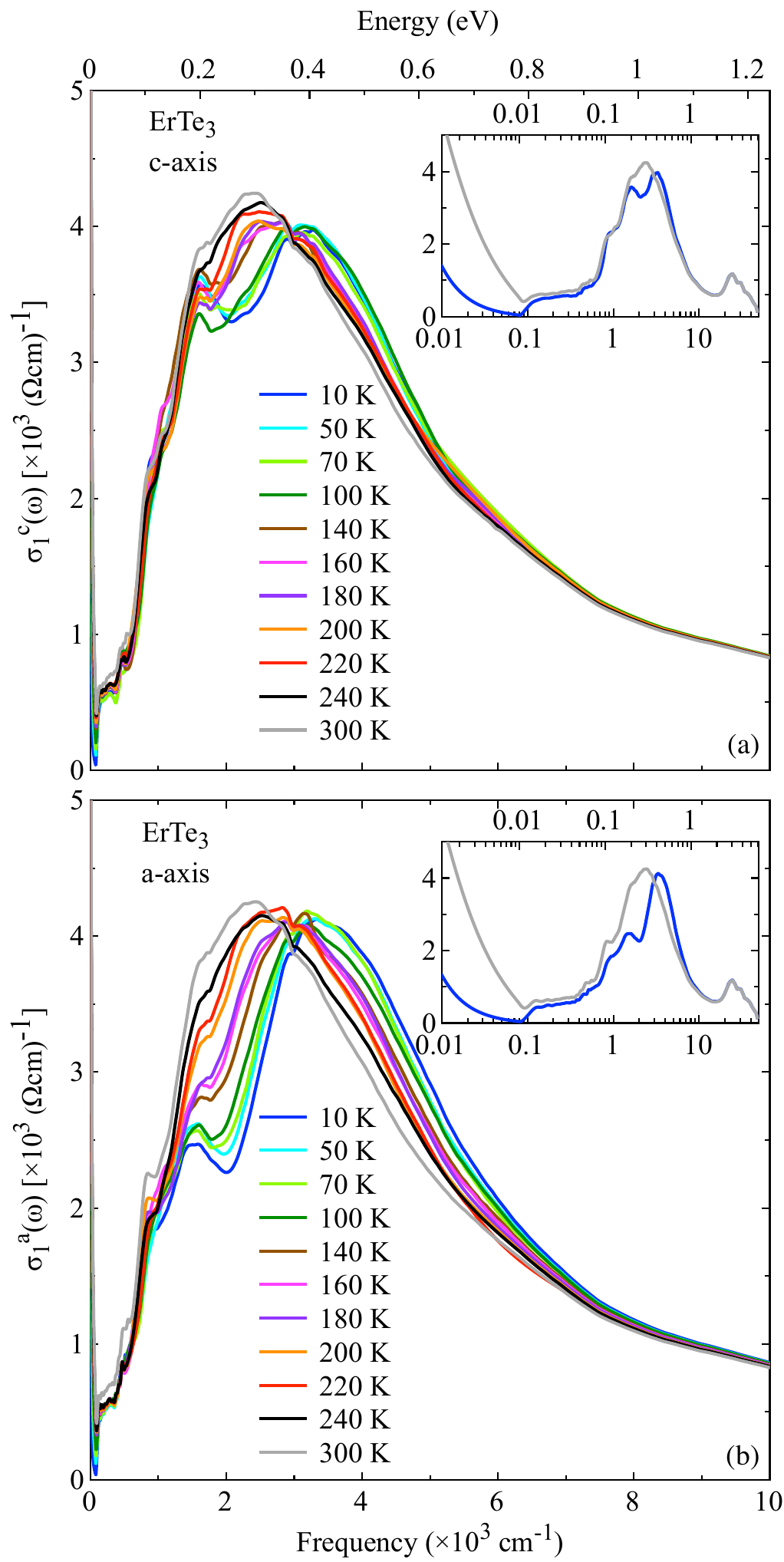}}
\caption{$T$ dependence of $\sigma_1(\omega)$ along the $c$ (a) and $a$ (b) axes (Fig. \ref{resistivity}(a)) of ErTe$_3$ (1 eV = 8.06548$\times$10$^3$ cm$^{-1}$), emphasising the MIR-NIR spectral range. The insets show the corresponding spectra at 300 and 10 K over the whole investigated spectral range (please note the logarithmic frequency/energy scale). 
}
\label{opt_cond_0}
\end{figure}

\begin{figure}
\centerline{\includegraphics[width=1\columnwidth]{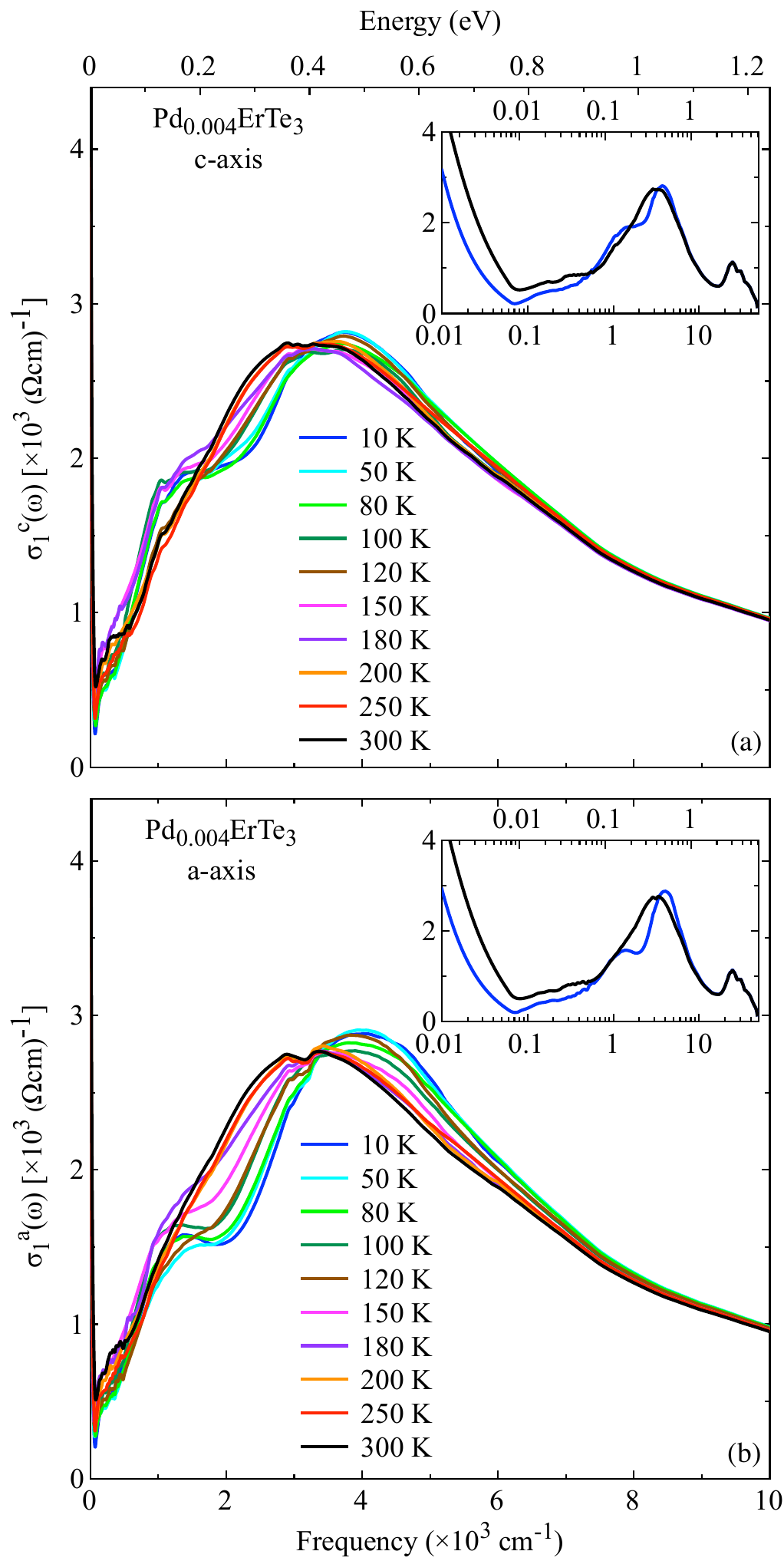}}
\caption{$T$ dependence of $\sigma_1(\omega)$ along the $c$ (a) and $a$ (b) axes (Fig. \ref{resistivity}(a)) of $x$ = 0.004 Pd$_x$ErTe$_3$ (1 eV = 8.06548$\times$10$^3$ cm$^{-1}$), emphasising the MIR-NIR spectral range. The insets show the corresponding spectra at 300 and 10 K over the whole investigated spectral range (please note the logarithmic frequency/energy scale).
}
\label{opt_cond_004}
\end{figure}

\begin{figure}%
\centerline{\includegraphics[width=1\columnwidth]{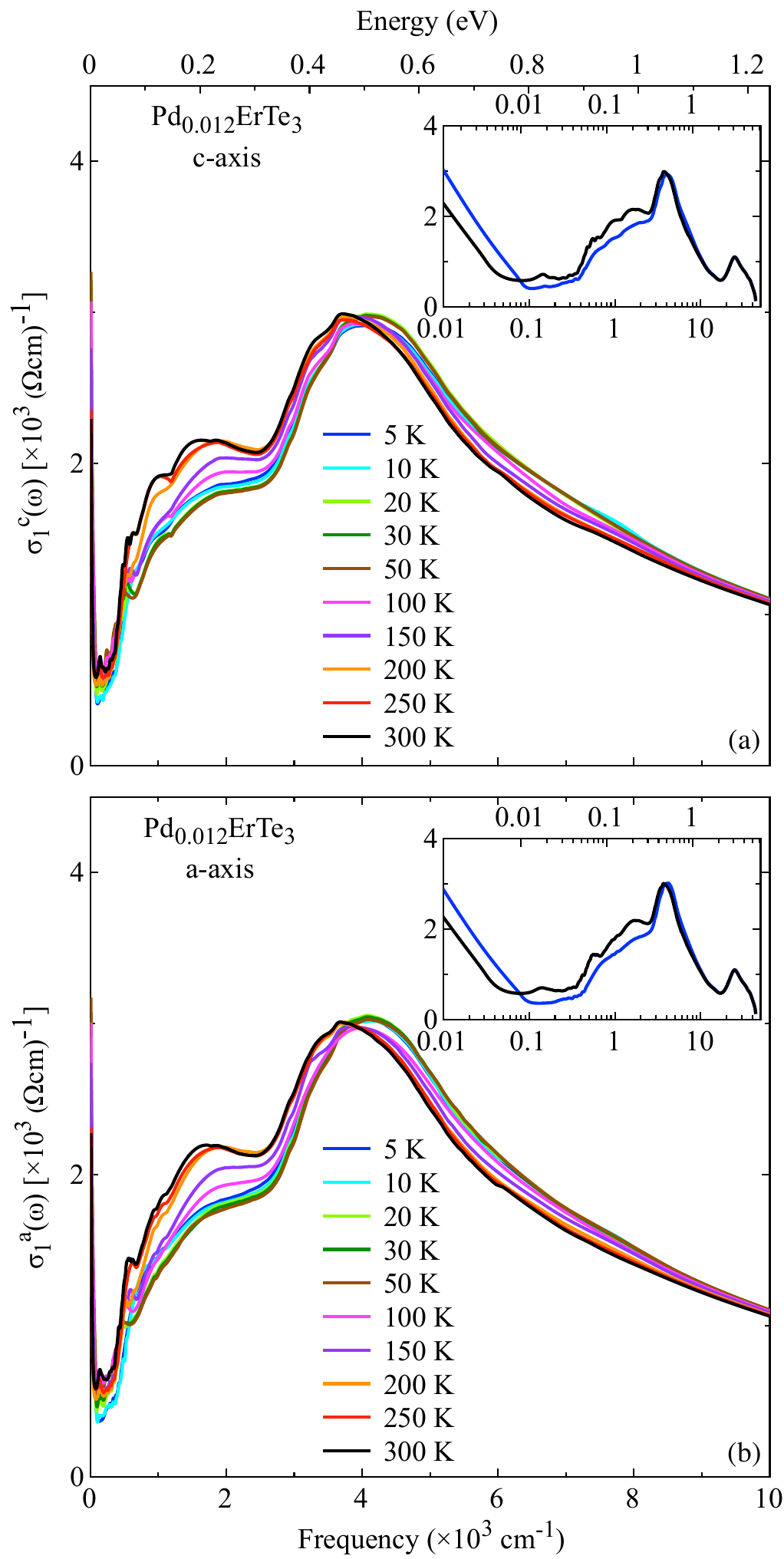}}
\caption{$T$ dependence of $\sigma_1(\omega)$ along the $c$ (a) and $a$ (b) axes (Fig. \ref{resistivity}(a)) of $x$ = 0.012 Pd$_x$ErTe$_3$ (1 eV = 8.06548$\times$10$^3$ cm$^{-1}$), emphasising the MIR-NIR spectral range. The insets show the corresponding spectra at 300 and 5 K over the whole investigated spectral range (please note the logarithmic frequency/energy scale).
}
\label{opt_cond_012}
\end{figure}

Figures \ref{opt_cond_0}-\ref{opt_cond_012} collect the derived $T$ dependence of the in-plane $\sigma_1(\omega)$, highlighting the spectral range below 10$^4$ cm$^{-1}$. The optical anisotropy between the $c$ and $a$ axes and its intercalation dependence as anticipated above for $R(\omega)$ are equally reproduced in $\sigma_1(\omega)$. As also shown in the corresponding insets, there are two relevant features in the spectra; a zero-energy mode, ascribed to the metallic Drude resonance, and a strong MIR absorption. The metallic component gets narrow with decreasing $T$ at low Pd-intercalations but broadens substantially for $x$ = 0.012 Pd$_x$ErTe$_3$. The Drude resonance and the MIR absorption are well distinct at all $T$ and may be also recognised in earlier work \cite{Pfuner2010,Hu2011}, yet more blurred upon increasing $T$. In fact, the MIR absorption was so far identified as a very broad feature at high $T$, almost disappearing into the large high frequency Drude tail \cite{Pfuner2010}. The peculiar $T$ dependence of the strong MIR absorption is then of interest. As emphasised by the main panels in Fig. \ref{opt_cond_0}, that MIR absorption in ErTe$_3$ consists of a peak with a shoulder on its low frequency side at high $T$, crossing over into two distinct peaks at low $T$. Such a behaviour is still observable in the $x$ = 0.004 Pd-intercalation (Fig. \ref{opt_cond_004}), while the low frequency shoulder of the MIR peak for the $x$ = 0.012 Pd-intercalation remarkably broadens upon lowering $T$ (Fig. \ref{opt_cond_012}).

A quite central quantity in our discussion is the spectral weight ($SW$) and its distribution \cite{SW}. $SW$ is generically defined through the direct integration of $\sigma_1(\omega)$ in chosen energy intervals between $\omega_i$ ($i$ = 1 and 2) so that $SW(T) = \frac{Z_0}{\pi^2}\int_{\omega_1}^{\omega_2}\sigma_1(\omega'; T)d\omega'$ (i.e., expressed in units of cm$^{-2}$ and with $Z_0$ = 376.73 $\Omega$ being the impedance of free space). This model-independent quantity is related to the number of the effective carriers (normalized by their effective mass) contributing to the optical processes within the integration limits and images the evolution of the electronic band structure upon varying $T$. In the case $\omega_1$ = 0 and for the $\omega_2\rightarrow\infty$ limit, the resulting integrated $SW(T)$ is expected to merge to a constant value at all \textit{T}, satisfying the optical $f$-sum rule \cite{Dressel2002}. The integrated $SW \sim \int_{0}^{\omega_c}\sigma_1(\omega)d\omega$ is here fully recovered at energy scales ($\omega_c$) above 10$^4$ cm$^{-1}$ and its reshuffling mainly occurs at MIR frequencies at all $T$. Before going any further, it is worth recalling, that if there is a transfer of $SW$ from high to low energies, the $SW$ ratio (i.e., $SW^{c,a}(T)/SW^{c,a}(300$ K), shown in Fig. 3) will exceed 1 at low energies and then smoothly approach 1 upon increasing frequency until the full energy scale of the low-energy resonance is reached. For instance, $SW$ may move into the metallic (Drude) zero-energy mode. If there is a transfer of $SW$ from low to high energies though, the $SW$ ratio will fall below 1 until the energy scale of the total $SW$ transfer is accomplished. This latter case may suggest some erosion of the density-of-states and a FS-gapping, as it could step in by a reconstruction of the electronic band structure.

\begin{figure}[h]
\centerline{\includegraphics[width=1\columnwidth]{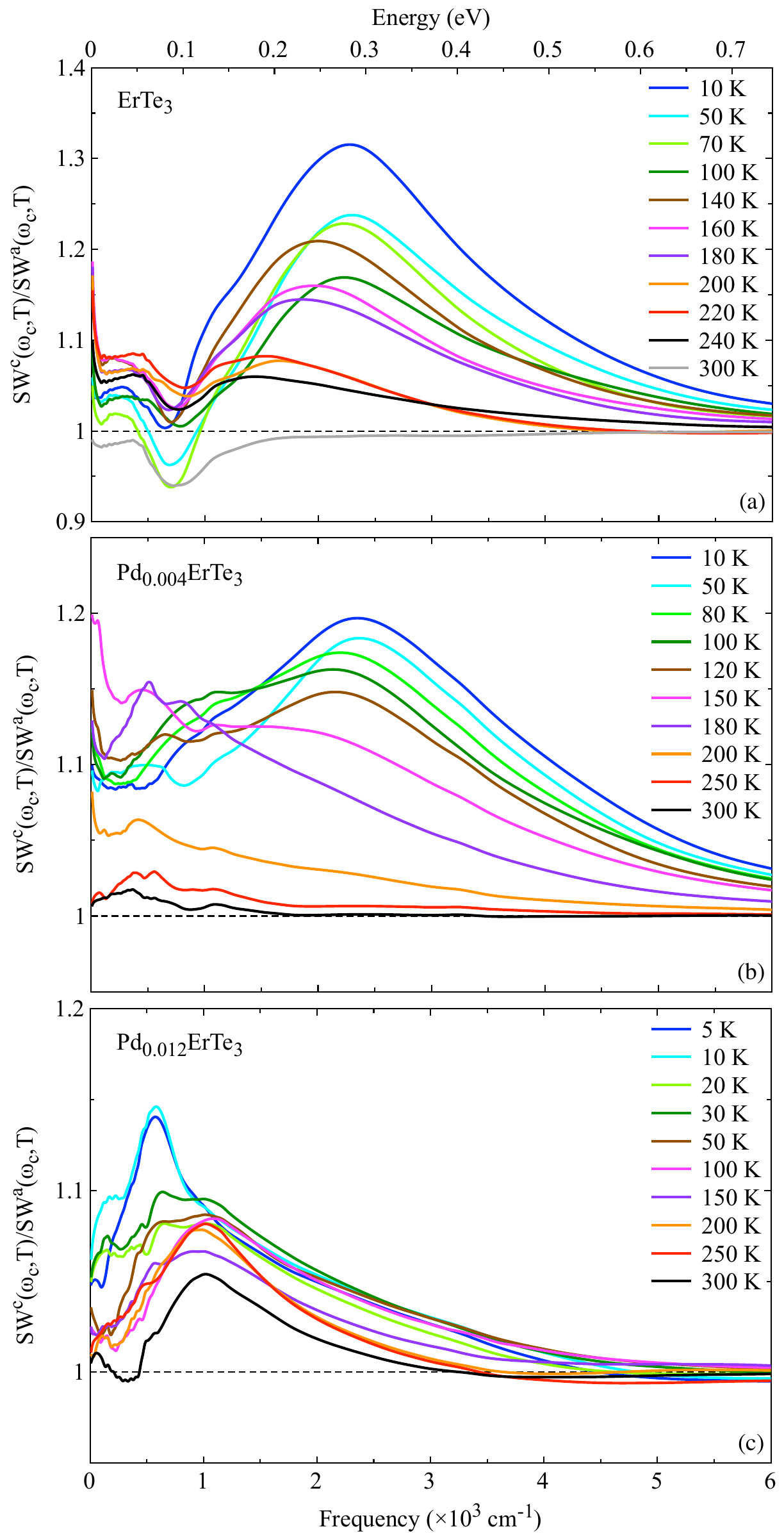}}
\caption{$T$ dependence of the ratio $SW^{c}(\omega_c, T)/SW^{a}(\omega_c, T)$ for the integrated $SW$ (insets Fig. 3) as a function of the cut-off frequency $\omega_c$ (see text) at selected $T$ for $x$ = 0 (a), $x$ = 0.004 (b) and $x$ = 0.012 (c) Pd$_x$ErTe$_3$ (1 eV = 8.06548$\times$10$^3$ cm$^{-1}$).
}
\label{int_SW}
\end{figure}

An alternative avenue is to focus on the $SW$ anisotropy. Figure \ref{int_SW} indeed displays the ratio $SW^{c}(\omega_c, T)/SW^{a}(\omega_c, T)$ of the integrated spectral weight up to $\omega_c$ between the two relevant crystallographic axes, denoting an obvious deviation from the isotropic limit (i.e., $\sim$ 1). That anisotropy (shown at selected $\omega_c$ in Fig. 2) smoothly evolves with lowering $T$ but overall decreases in magnitude with Pd-intercalation. As evinced from other quantities (like e.g. $\sigma_1(\omega)$), the affected spectral range tends to shift towards low energy scales, concerning the FIR-MIR energy interval, upon Pd-intercalation. All that testifies an anisotropic electronic band structure across and below $E_F$ in coincidence with the CDW transitions.

\subsection{Phenomenological data analysis}

We fit the $T$ dependence of the optical response using the phenomenological Drude-Lorentz approach, for which the dielectric function $\tilde{\varepsilon}=\varepsilon_{1}+i\varepsilon_{2}$ is given by \cite{Dressel2002}:
%
%
\begin{equation}
\tilde{\varepsilon}(\omega)=\varepsilon_{\infty}-\frac{\omega^{2}_{p,D}}
 {\omega^{2}+\frac{i\omega}{\tau_{D}}}+\sum_{j}\frac{\Omega^{2}_{j}}{\omega^2_{j}-\omega^2-i\omega\gamma_{j}},
\label{fit}
\end{equation}
$\varepsilon_{\infty}$ is the optical dielectric constant, which turns out in our fits to be close to 1 for all Pd$_x$ErTe$_3$.
In the second term, $\Gamma_{D} = 1/\tau_{D}$ is the scattering rate and $\omega^{2}_{p,D}= 4\pi ne^{2}/m^{*}$ is the squared plasma frequency for the itinerant (Drude) carriers, with $n$ and $m^{*}$ as the carriers concentration and effective mass, respectively. $\omega^{2}_{p,D}$ equally defines the so-called Drude weight.
In the third term, $\omega_{j}$, $\gamma_{j}$ and $\Omega_{j}$ are the resonance frequency, width and strength, respectively, of the $j$th Lorentz (Lj) harmonic oscillators (HO), which describe bound excitations. The square of their strength ($\Omega_{j}^2$) corresponds to $SW$ encountered in each Lj excitation. The use of the Drude and Lj-HOs $SW$ is obviously an alternative approach to the direct integration of the measured quantity $\sigma_1(\omega)$, as introduced above and elaborated in the main text. We will return however below to a comparison between the two methods, pointing then out their mutual consistency and complementarity. 

\begin{figure}[h]
\centerline{\includegraphics[width=1\columnwidth]{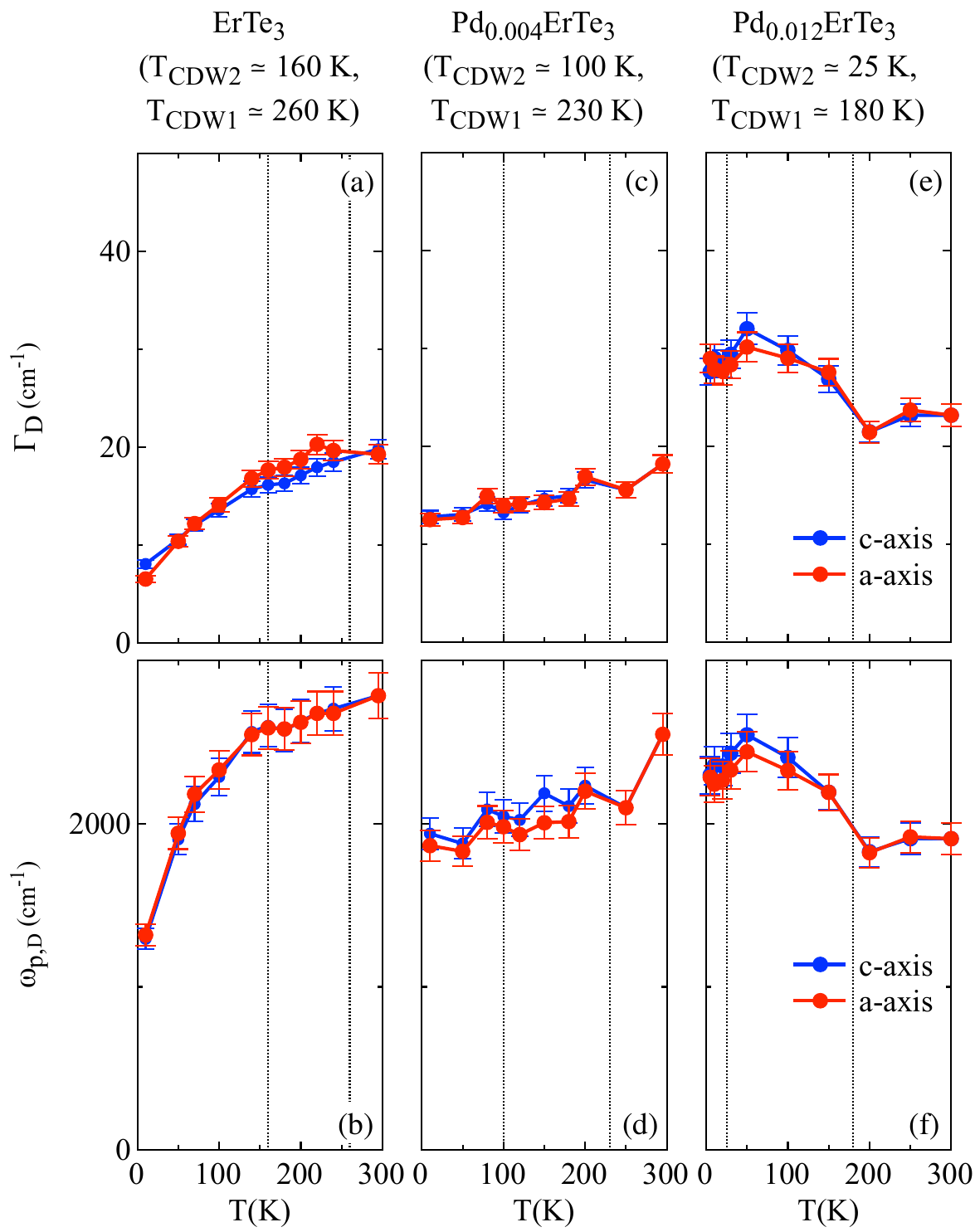}}
\caption{$T$ dependence of the Drude scattering rate $\Gamma_D$ (upper panels) and plasma frequency $\omega_{p,D}$ (lower panels) fit parameters (Eq. \ref{fit}) for all Pd$_x$ErTe$_3$ along the $c$ and $a$ axes: (a,b) $x$ = 0, (c,d) $x$ = 0.004 and (e,f) $x$ = 0.012. The vertical dotted lines mark $T_{CDW1}$ and $T_{CDW2}$ \cite{Walmsley2017,Straquadine2019}. The error bars are estimated numerically within the non-linear least-squares fit technique. 
}
\label{Drude_param}
\end{figure}

\begin{figure}[h]
\centerline{\includegraphics[width=1\columnwidth]{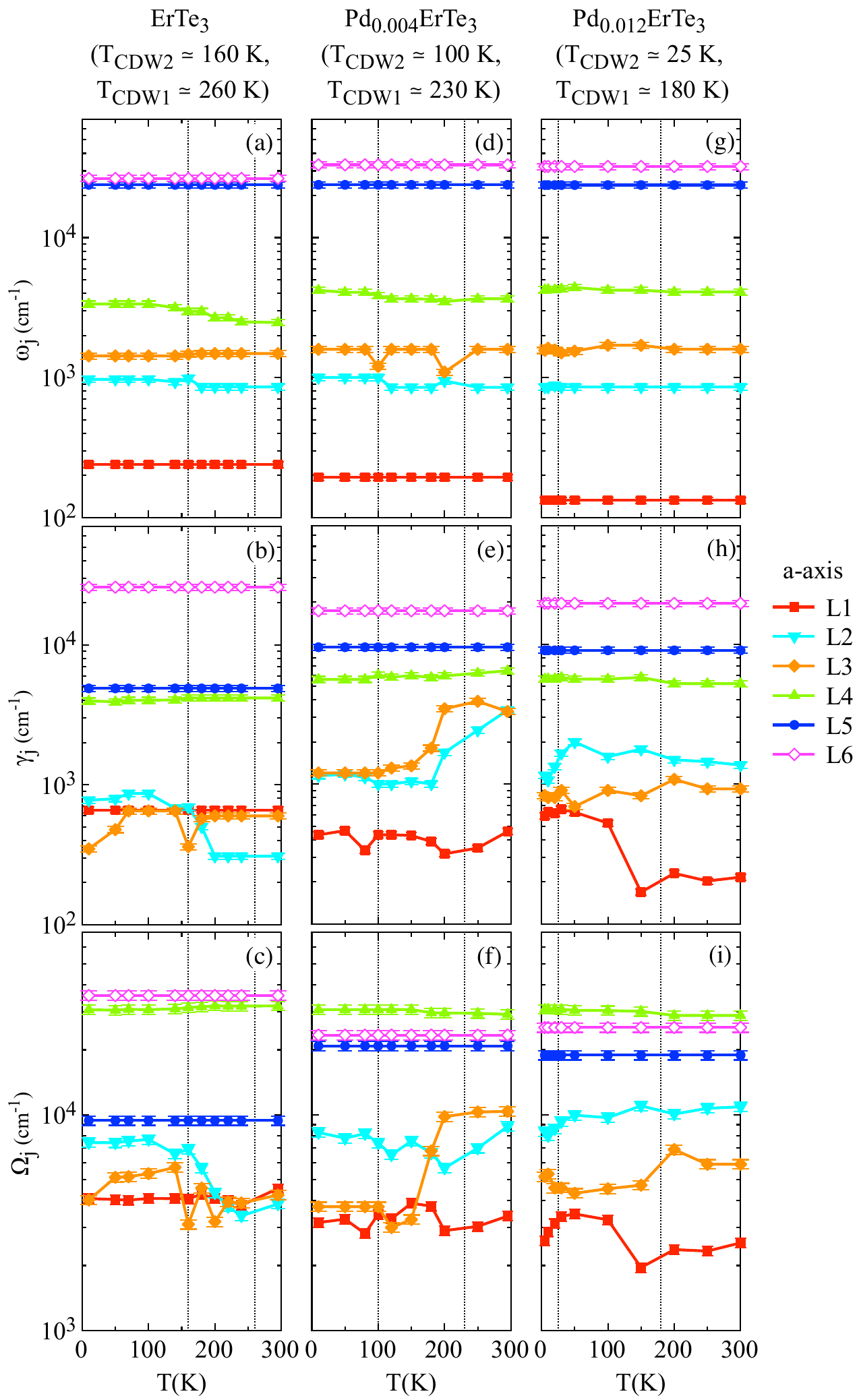}}
\caption{$T$ dependence of the Lj-HOs resonance frequency $\omega_j$ (upper panels), damping $\gamma_j$ (middle panels) and strength $\Omega_j$ (lower panels) fit parameters (Eq. \ref{fit}) for all Pd$_x$ErTe$_3$ along the $a$ axis: (a-c) $x$ = 0, (d-f) $x$ = 0.004 and (g-i) $x$ = 0.012. The vertical dotted lines mark $T_{CDW1}$ and $T_{CDW2}$ \cite{Walmsley2017,Straquadine2019}. The error bars are estimated numerically within the non-linear least-squares fit technique. 
}
\label{HO_param_0}
\end{figure}

\begin{figure}[h]
\centerline{\includegraphics[width=1\columnwidth]{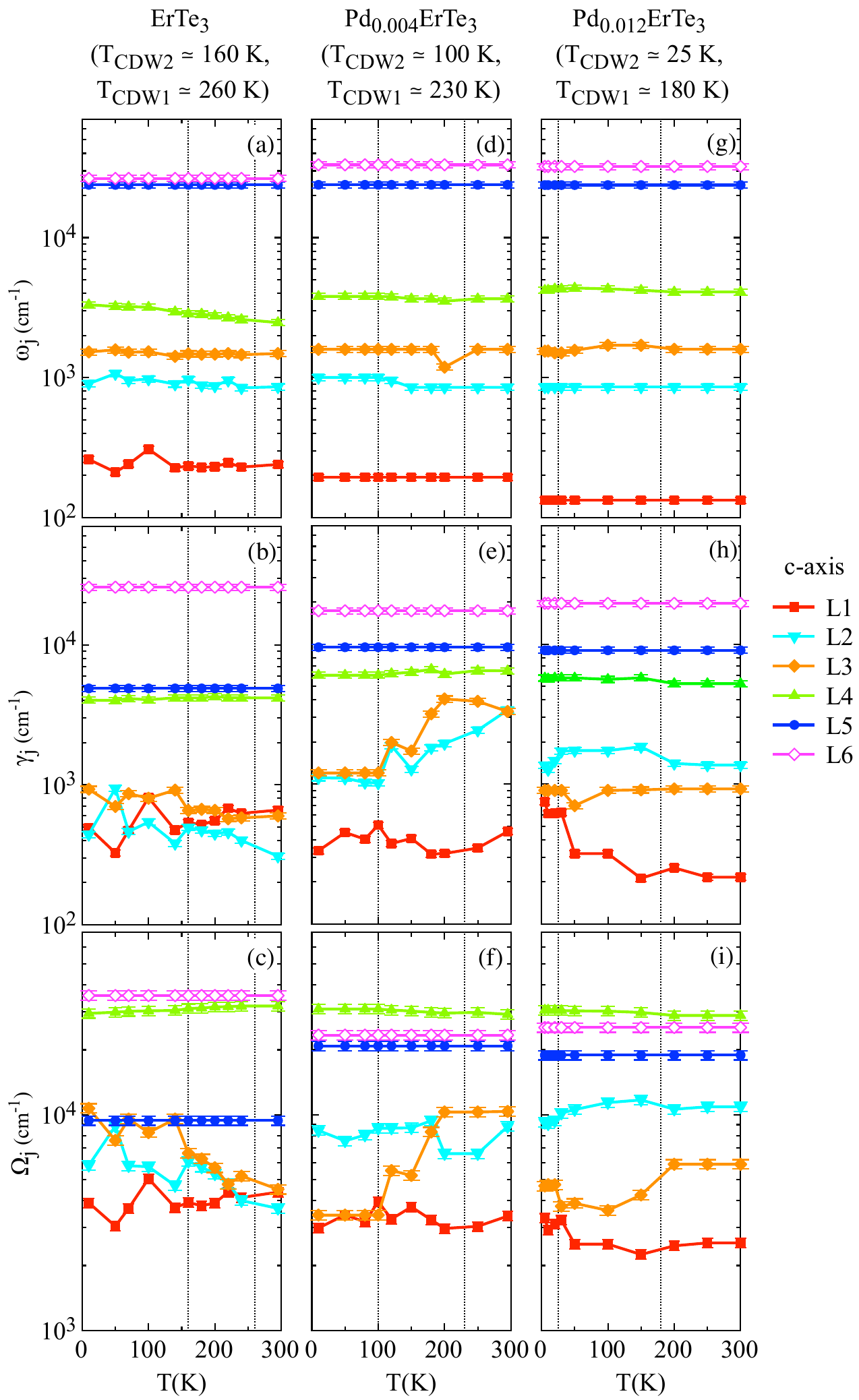}}
\caption{$T$ dependence of the Lj-HOs resonance frequency $\omega_j$ (upper panels), damping $\gamma_j$ (middle panels) and strength $\Omega_j$ (lower panels) fit parameters (Eq. \ref{fit}) for all Pd$_x$ErTe$_3$ along the $c$ axis: (a-c) $x$ = 0, (d-f) $x$ = 0.004 and (g-i) $x$ = 0.012. The vertical dotted lines mark $T_{CDW1}$ and $T_{CDW2}$ \cite{Walmsley2017,Straquadine2019}. The error bars are estimated numerically within the non-linear least-squares fit technique. 
}
\label{HO_param_90}
\end{figure}

\begin{figure}[h]
\centerline{\includegraphics[width=1\columnwidth]{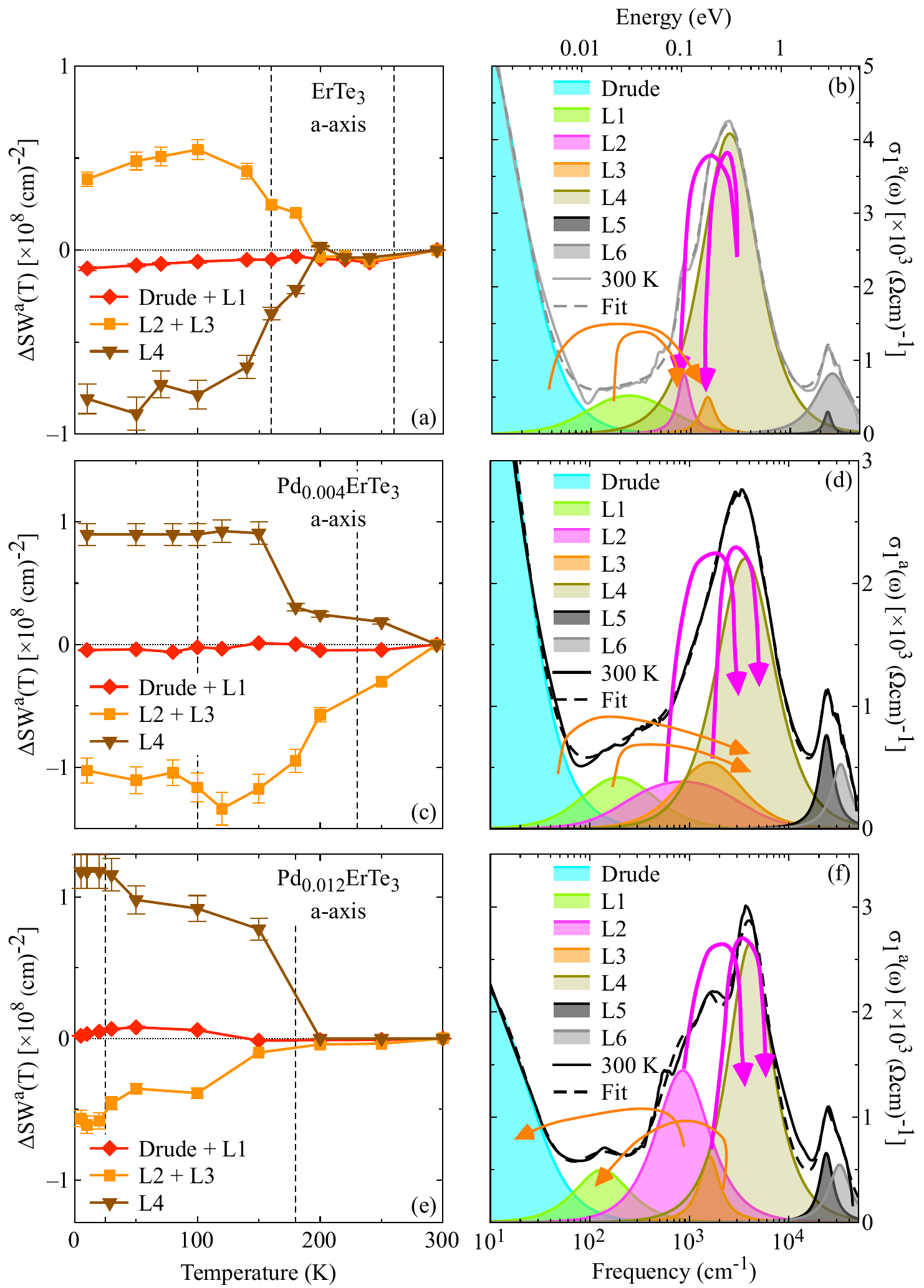}}
\caption{Left column: $T$ dependence of the spectral weight relative variation $\Delta SW(T) = SW(T) - SW$(300 K), calculated for the grouped contributions given by the Drude term and HO L1 (i.e., $SW \sim \omega_{p,D}^2 + \Omega_{1}^2$), by HOs L2 and L3 (i.e., $SW \sim \Omega_{2}^2 + \Omega_{3}^2$), as well as by HO L4 (i.e., $SW \sim \Omega_{4}^2$) with respect to 300 K. The error bars in $\Delta SW(T)$ correspond to the direct propagation of the error in the HOs strength, estimated numerically within the non-linear least-squares fit technique. Right column: $\sigma_1(\omega)$ at 300 K compared to its Drude-Lorentz fit (Eq. \ref{fit}), showing its single constituent components (please note the logarithmic frequency/energy scale). The rounded arrows mimic the direction of the $SW$ reshuffling upon decreasing $T$, which is stronger with thicker arrows. All quantities are shown along the $a$ axis (Fig. \ref{resistivity}(a)) for all Pd$_x$ErTe$_3$: (a-b) $x$ = 0, (c-d) $x$ = 0.004 and (e-f) $x$ = 0.012. The vertical dashed lines in panels (a,c,e) mark $T_{CDW1}$ and $T_{CDW2}$ \cite{Walmsley2017,Straquadine2019}.
}
\label{SW_fit0}
\end{figure}

\begin{figure}[h]
\centerline{\includegraphics[width=1\columnwidth]{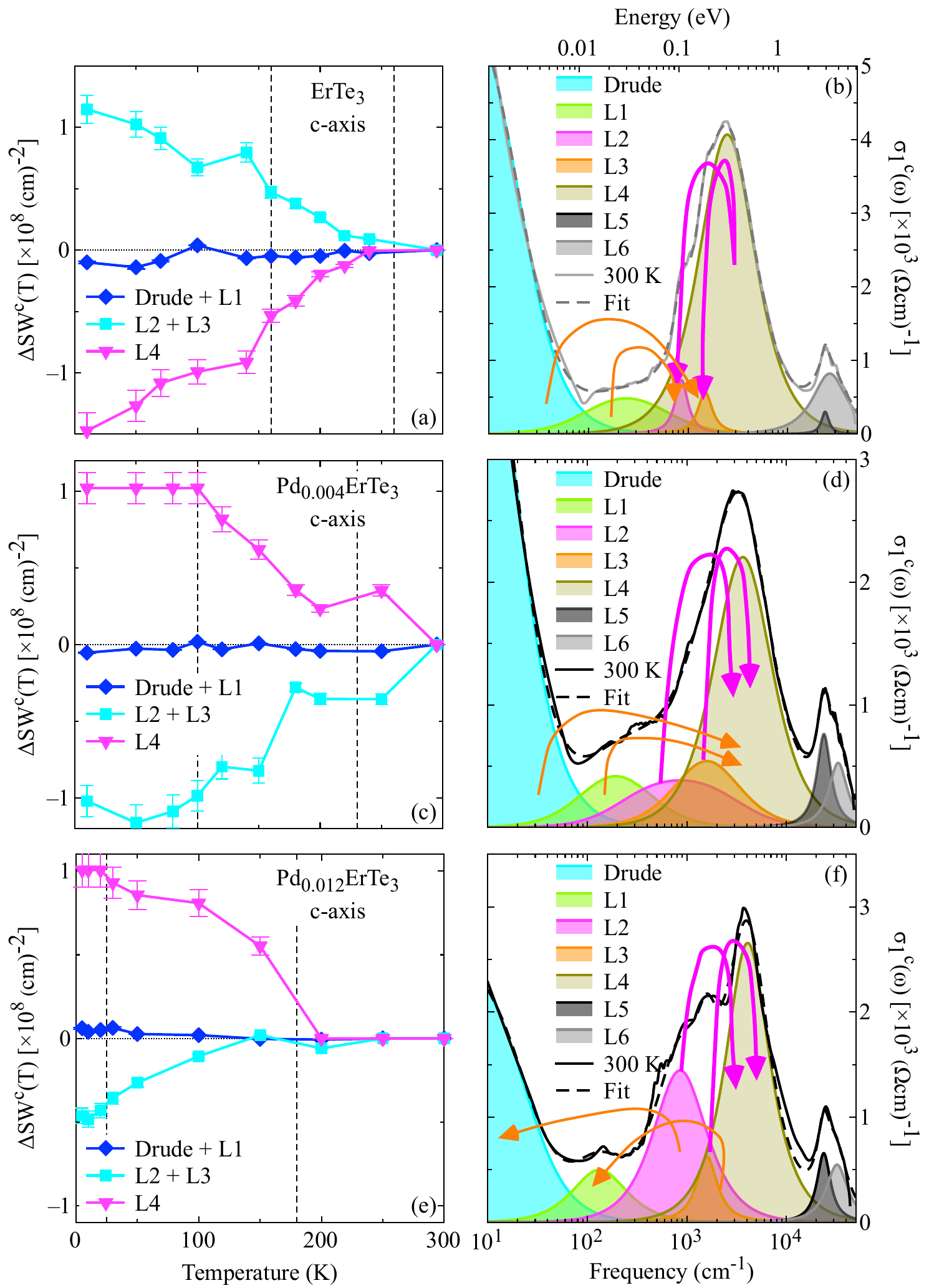}}
\caption{Left column: $T$ dependence of the spectral weight relative variation $\Delta SW(T) = SW(T) - SW$(300 K), calculated for the grouped contributions given by the Drude term and HO L1 (i.e., $SW \sim \omega_{p,D}^2 + \Omega_{1}^2$), by HOs L2 and L3 (i.e., $SW \sim \Omega_{2}^2 + \Omega_{3}^2$), as well as by HO L4 (i.e., $SW \sim \Omega_{4}^2$) with respect to 300 K. The error bars in $\Delta SW(T)$ correspond to the direct propagation of the error in the HOs strength, estimated numerically within the non-linear least-squares fit technique. Right column: $\sigma_1(\omega)$ at 300 K compared to its Drude-Lorentz fit (Eq. \ref{fit}), showing its single constituent components (please note the logarithmic frequency/energy scale). The rounded arrows mimic the direction of the $SW$ reshuffling upon decreasing $T$, which is stronger with thicker arrows. All quantities are shown along the $c$ axis (Fig. \ref{resistivity}(a)) for all Pd$_x$ErTe$_3$: (a-b) $x$ = 0, (c-d) $x$ = 0.004 and (e-f) $x$ = 0.012. The vertical dashed lines in panels (a,c,e) mark $T_{CDW1}$ and $T_{CDW2}$ \cite{Walmsley2017,Straquadine2019}.
}
\label{SW_fit90}
\end{figure}

The real part $\sigma_1(\omega)$ of the optical conductivity is specifically at the centre of our analysis, for which we seek a series of additive components being equivalent for all Pd-intercalations and for both polarisation directions. It turns out that a reliable fit covering the whole measured spectral range consists of one Drude term and six HOs. Figures \ref{Drude_param}, \ref{HO_param_0} and \ref{HO_param_90} display the $T$ dependence of the Drude as well as of the Lj-HOs parameters, while the panels of the right column in Figs. \ref{SW_fit0} and \ref{SW_fit90} visualise the fit components across the spectral range of interest along both the $a$ and $c$ axes. The same fit philosophy, which copies rather well with a previous analysis nonetheless for unpolarised data on ErTe$_3$ \cite{Pfuner2010}, allows a fairly precise reproduction of the spectra at any $T$ and polarisation directions for all Pd-intercalations, thus supporting a comprehensive and consequent discussion even beyond the intrinsic model-dependent nature of this phenomenological approach.

We first comment on the fit parameters of the Drude term. Both the $x$ = 0 and 0.004 Pd-intercalations display a monotonous decrease of the scattering rates and plasma frequencies with decreasing $T$ (Figs. \ref{Drude_param}(a-d)). The $T$ dependence of $\Gamma_{D}$ indicates a typical, effective metallic behaviour, while that of $\omega_{p,D}$ suggests an erosion of the metallic $SW$ which gets quite pronounced for $x$ = 0 at low $T$. These trends are again an indication for a progressively enhanced gapping of FS because of the CDW transitions (see an alternative elaboration around Fig. 3), similarly accompanied by a reduction of available scattering channels. Somehow opposite is the situation for the $x$ = 0.012 Pd-intercalation, since both Drude parameters increase upon lowering $T$ below $T_{CDW1}$ prior being weakly depleted at the lowest measured $T$ below $T_{CDW2}$ (Figs. \ref{Drude_param}(e,f)). Therefore, only at the lowest $T$ an effective but still partial gapping of FS and a modest switching-off of scattering channels can be optically mirrored as a consequence of the CDW transitions. Otherwise, disorder by Pd-intercalation seems here to dominate at $T$ between the two CDW transitions, with increasing scattering. This is then compensated by an increasing Drude plasma frequency for $T_{CDW2} < T < T_{CDW1}$, denoting some enhancement of FS with more itinerant, yet less mobile charge carriers due to impurity bands. All in all, this may presage a nearly saturated $T$ dependence of $\rho_{dc}(T)$ for both in-plane axes besides the broad, humble anomaly below $T_{CDW1}$, as observed for instance in the $x$ = 0.019 composition (Fig. \ref{resistivity}(d)). At this stage, it is worth warning the readership that it is admittedly challenging to perform a unique as well as fully reliable fit of the zero-energy resonance, since such an effective metallic component of $\sigma_1(\omega)$ is very narrow and at low $T$ even quite resolution-limited (i.e., almost fully beyond the measured experimental spectral range). Therefore, despite the 
fair qualitative agreement between the measured $dc$ transport properties and their anisotropy and the calculated one from the Drude model, there is no pretension about any quantitative conclusions. However, as addressed below in terms of the encountered $SW$, the metallic spectral range in $\sigma_1(\omega)$ plays a marginal role with respect to the FIR-MIR energy interval.

As far as the fit parameters of HOs are concerned, we remark an almost monotonous $T$ dependence, with some exceptions (Figs. \ref{HO_param_0} and \ref{HO_param_90}): HOs L5 and L6 turn out to be fully $T$ independent and specifically HOs L1 to L3 show some abrupt or sudden changes in $T$, particularly for their damping and strength, when crossing either CDW transitions. This means that the CDW transitions do not only affect the bands crossing $E_F$ and consequently the Drude fit parameters (Fig. \ref{Drude_param}) but also those bands deep into the electronic structure (i.e., far from $E_F$), suggesting its (intercalation-driven) remodelling. This is addressed next from the perspective of the $SW$ redistribution.

Complementary to our analysis of the integrated $SW$ in the main text, we can now single out the relevant spectral ranges harbouring the $T$-dependent $SW$ reshuffling. The fit components themselves, for which $SW \sim \omega_{p,D}^2$ (Drude) and $SW \sim \Omega_{j}^2$ (Lorentz Lj-HO), identify the selected spectral ranges. We state the redistribution of $SW$ as its relative change with respect to 300 K, i.e. $\Delta SW(T) = SW(T) - SW$(300 K) (Figs. \ref{SW_fit0} and \ref{SW_fit90} for the $a$ and $c$ axes, respectively). For the $x$ = 0 and 0.004 Pd-intercalations and along both axes (Figs. \ref{SW_fit0}(a,c) and \ref{SW_fit90}(a,c)), there is a weak depletion of $SW$ upon lowering $T$ related to the Drude term (as anticipated above when presenting Figs. \ref{Drude_param}(b,d)) and its high frequency tail incorporated by HO L1 (i.e., Drude + L1 in Figs. \ref{SW_fit0}(a,c) and \ref{SW_fit90}(a,c) refers to $SW \sim \omega_{p,D}^2 + \Omega_{1}^2$). The $x$ = 0.012 Pd-intercalation is though at variance. As already said in relation to Fig. \ref{Drude_param}(f) and for both axes, some very moderate $SW$ first moves to the Drude term and its tail represented by HO L1 with decreasing $T$ but then below $T_{CDW2}$ a renewed and weak $SW$ suppression takes place at the metallic components. It is striking that the damping and partially the strength of HO L1 for the $x$ = 0.012 Pd-intercalation are steadily enhanced for $T_{CDW2} < T < T_{CDW1}$, which could also mimic disorder-driven localisation effects of the itinerant charge carriers \cite{Huang2012}, prior the onset of the CDW condensate at $T < T_{CDW2}$.

The moderate $SW$ redistribution at low energy scales for all Pd-intercalations is accompanied by the stronger $SW$ reshuffling pertinent to the spectral range covered by HOs L2 and L3 (i.e., L2 + L3 in Figs. \ref{SW_fit0}(a,c,e) and \ref{SW_fit90}(a,c,e) refers to $SW \sim \Omega_{2}^2 + \Omega_{3}^2$) as well as L4 (i.e., $SW \sim \Omega_{4}^2$), and being similar along both axes. For $x$ = 0, $SW$ of HO L4 is removed and is then globally redistributed at low frequencies, accumulating into HOs L2 and L3 (Figs. \ref{SW_fit0}(a) and \ref{SW_fit90}(a)), when lowering $T$. Exactly the reversed behaviour is observed for the $x$ = 0.004 and 0.012 Pd-intercalations (Figs. \ref{SW_fit0}(c,e) and \ref{SW_fit90}(c,e)). In fact, $SW$ along both axes accumulates into HO L4 at the partial costs of $SW$ in HOs L2 and L3 at low $T$. Such trends upon lowering $T$ seem to be initially rather gradual with steady increase/depletion of $\Delta SW$ upon crossing $T_{CDW1}$ and lean to a saturation of $\Delta SW$ below $T_{CDW2}$ for all Pd-intercalations. Interestingly enough, for the $x$ = 0 and 0.004 Pd-intercalations the $SW$ redistribution fully occurs within the indicated energy ranges related to each Lj-HOs and Drude term, since the involved total $SW$ in the spectral range addressed here remains constant at any $T$ (i.e., $\sum_i\Delta SW_i \simeq$ 0, with $i$ running over each combinations of the grouped fit components in Figs. \ref{SW_fit0} and \ref{SW_fit90} for both the $a$ and $c$ axes). On the contrary, for the $x$ = 0.012 Pd-intercalation the $SW$ reshuffling apparently does not guarantee its conservation at low $T$ within the discussed spectral ranges for both axes; we may speculate that additional $SW$ should move with decreasing $T$ into the absorption described by HO L4 from excitations at higher energy scales and distributed over a large energy interval, eventually not accessible to our experiment.

Summarising, the rounded arrows in Figs. \ref{SW_fit0}(b,d,f) and \ref{SW_fit90}(b,d,f) pictorially highlight the direction and with increasing thickness catch a glimpse of the amount of the $SW$ reshuffling upon decreasing $T$. This representation features the overall equivalent $SW$ redistribution for both axes at each Pd-intercalation. The metallic part (Drude) and its incoherent tail (HO L1) loose $SW$ in favour of the low frequency shoulder (HOs L2 and L3) of the MIR peak for the $x$ = 0 and of the MIR peak itself (HO L4) for the $x$ = 0.004 Pd-intercalation with decreasing $T$, respectively. On the other hand, that low frequency shoulder (HOs L2 and L3) also profits and gets additionally reinforced from $SW$ of the MIR peak (HO L4) in $x$ = 0 but reversely looses $SW$ towards the MIR peak for the $x$ = 0.004 Pd-intercalation upon lowering $T$. For the $x$ = 0.012 Pd-intercalation, the low frequency shoulder (HOs L2 and L3) of the MIR peak experiences a diminishing $SW$ at low $T$ towards the metallic part of $\sigma_1(\omega)$ as well as towards the MIR peak (HO L4), contributing to its reinforcement. However, the major $SW$ reshuffling when crossing the CDW transitions for all Pd-intercalations effectively happens at and within the energy scales of the strong MIR peak in $\sigma_1(\omega)$, widely associated with the spectral range of the CDW gap(s) and spanned by the combination of the phenomenological HOs L2 to L4. Intriguingly enough, the $SW$ reallocation along both axes at low $T$ is towards the low frequency tail of the MIR peak (i.e., from HO L4 to HOs L2+L3) for the pristine materials, while it reinforces its high frequency absorption (i.e., from HOs L2+L3 to HO L4) as soon as Pd-intercalation takes place.

%

\end{document}